\documentclass[twocolumn]{aa}  

\usepackage{graphicx}
\usepackage{placeins}
\usepackage{txfonts}
\usepackage{textcomp}
\usepackage{multirow}
\usepackage{dcolumn}
\usepackage{xcolor}
\usepackage{amsmath,amssymb}
\usepackage[modulo, switch, mathlines]{lineno}
\modulolinenumbers[1]

\begin{document}

\title{Internal rotation of 13 low-mass low-luminosity red giants in the \emph{Kepler} field}

\author{S.~A. Triana \inst{\ref{inst1},\ref{inst6}}
          \and
          E. Corsaro \inst{\ref{inst2},\ref{inst3},\ref{inst4}}
          \and
          J. De Ridder \inst{\ref{inst1}}
          \and
          A. Bonanno \inst{\ref{inst5}}
          \and
          F. P\'erez Hern\'andez\inst{\ref{inst2},\ref{inst4}}
          \and
          R.~A. Garc\'ia \inst{\ref{inst3}}
          }
\authorrunning{Triana, Corsaro, et al.}
\titlerunning{Internal rotation of 13 red giants}

\institute{
                Institute of Astronomy, KU Leuven, Celestijnenlaan 200D, 3001 Leuven, Belgium
                \label{inst1}
                \and
                Instituto de Astrof\'isica de Canarias, E-38200 La Laguna, Tenerife, Spain
                \label{inst2}
                \and
                Laboratoire AIM, CEA/DRF-CNRS, Université Paris 7 Diderot, IRFU/SAp, Centre de Saclay, 91191 Gif-sur-Yvette, France
                \label{inst3}
                \and
                Departamento de Astrofísica, Universidad de La Laguna, E-38205 La Laguna, Tenerife, Spain
                \label{inst4}
                \and
                INAF, Osservatorio Astrofisico di Catania, Via S.Sofia 78, 95123 Catania,Italy
                \label{inst5}
                \and
                Royal Observatory of Belgium, Ringlaan 3, Brussels, Belgium
                \email{trianas@oma.be} \label{inst6}
          }

   \date{Received ; accepted }

 
  \abstract
   {The Kepler space telescope has provided time series of red giants of such unprecedented
 quality that a detailed asteroseismic analysis becomes possible. For a limited set of about
 a dozen red giants, the observed oscillation frequencies obtained by peak-bagging together
 with the most recent pulsation codes allowed us to reliably determine the core/envelope
 rotation ratio. The results so far show that the current models are unable to reproduce
 the rotation ratios, predicting higher values than what is observed and thus indicating
 that an efficient angular momentum transport mechanism should be at work. Here we provide
 an asteroseismic analysis of a sample of 13 low-luminosity low-mass red giant stars 
observed by \textit{Kepler} during its first nominal mission. These targets form a subsample 
of the 19 red giants studied previously \citet{corsaro2015}, which not only have
 a large number of extracted oscillation frequencies, but also unambiguous mode identifications.}
   {We aim to extend the sample of red giants for which internal rotation ratios obtained
 by theoretical modeling of peak-bagged frequencies are available. We also derive the rotation ratios
 using different methods, and compare the results of these methods with each other.}
   {We built seismic models using a grid search combined with a Nelder-Mead simplex
    algorithm and obtained rotation averages employing Bayesian inference and inversion methods. We compared these
    averages with those obtained using a previously developed
model-independent method.}
   {We find that the cores of the red giants in this sample are rotating 5 to 10 times faster than their
    envelopes, which is consistent with earlier results.
   The rotation rates computed from the different methods show good agreement for some targets, while
   some discrepancies exist for others.}
   {}


   \maketitle
%

\section{Introduction}

The impact of the \textit{Kepler} space mission \citep{borucki2010,koch2010} on diverse aspects of stellar astrophysics
has been enormous and revolutionary by many standards. Our understanding
of stellar evolution through asteroseismology has improved dramatically, 
and with this improvement, new challenges have appeared.

Sun-like stars, particularly red giants, exhibit 
a very rich pulsation pattern \citep{deridder2009, stello2009, hekker2009}. Some of these
pulsations can be associated with pressure ({\it p}) modes,
which are excited stochastically by turbulent convection.
These {\it p} modes propagate throughout the star with the highest sensitivity
 to the external convective envelope, as opposed to the internal gravity
 ({\it g}) modes, which propagate only throughout the radiative core and hence
are beyond observational reach. 
The {\it p} and {\it g} propagation zones generally do not overlap, and the region between
them is called the evanescent zone. Modes of mixed character, behaving like {\it g} modes
in the core and {\it p} modes in the envelope, bridge the evanescent zone
and have substantial amplitudes in both the core and the envelope
\citep{beck2011}. These modes are extremely useful for obtaining information
about the internal rotation of the core. This is possible because
rotation induces frequency splittings in the modes, which would otherwise be
degenerate in a spherically symmetric star \citep{ledoux1951}. Rotation induces a preferential
axis in the star, and if the rotation rate is much smaller than
the pulsation frequencies, then the splitting $\delta_{nlm}$ of a mode with radial,
angular, and azimuthal wavenumbers $n,l,m$ can be computed as \citep[see, e.g.,][\,]{Aerts2010asteroseismology}
\begin{linenomath*}
\begin{equation}
\label{forward}
\delta_{nlm}=m\, \beta_{nl} \int_0^{R_*} K_{nl}(r)\, \Omega(r)\, \mathrm{d}r,
\end{equation}
\end{linenomath*}
where the {\it kernel} $K_{nl}(r)$ and $\beta_{nl}$ are functions of the
vertical and horizontal material displacement eigefunctions $\xi_r(r)$ and $\xi_h(r)$ 
(see Section\,3 for more details).
Thus, rotation lifts the azimuthal wave number degeneracy of the modes.

The milestone work of \citet{beck2012} on red giants, using data obtained
by the \textit{Kepler }space telescope, took advantage of the detection of 
rotationally split mixed modes 
in KIC\,8366239 and concluded that the stellar core spins about ten times faster
than the envelope. In the same year, \citet{mosser2012} presented the core rotation of a
sample of 300 red giants, establishing that the cores slow down significantly during
the last stages of the red giant branch. A number of other studies followed. Most notably,
 \citet{deheuvels2012} determined a core/envelope rotation rate ratio of about
five for a star in the lower giant branch observed by \textit{Kepler}, 
\citet{deheuvels2014} computed the rotation rate ratio of six subgiants and young
red giants, \citet{deheuvels2015} obtained rotations rates for seven
red giants in the secondary clump, and very recently, the work by \citet{dimauro2016} resolved
the core rotation better than previous studies. In all these cases, the rotation rate
of the core does not match the expectations of current angular momentum theories.
Indeed, according to our current understanding of the evolution
of the angular momentum in stellar interiors, the core is expected
to spin up considerably as it contracts in stars at this stage of
evolution. \citet{cantiello2014} explicitly showed the inadequacy of current models in reproducing
the observed slow rotation of the core in RGB stars, even after magnetic
effects were included. It is then clear that a very effective mechanism for angular
momentum transport is at work. Internal gravity waves are capable of transferring
considerable amounts of angular momentum \citep{rogers2015,alvan2013}, which
provides a suitable explanation for `anomalous' rotation rates in other types of stars,
 such as those reported by \citet{kurtz2014,saio2015}, and \citet{triana2015}.
However, \citet{fuller2014} showed that this mechanism falls short
 of explaining the observed rotation rate ratios in stars on the red giant branch.
 However, mixed modes can also transport angular momentum, as demonstrated recently by \citet{
 belkacem2015}. According to this study, the mixed-mode wave heat flux has an appreciable
effect on the mean angular momentum in the inner regions of the star.

The methods used to obtain the internal rotation rates in red giant stars have seen
a number of developments well worth mentioning here. The way to proceed,
after the mode detection and identification process \citep[see, e.g.,][]{corsaro2015} 
is usually to develop a seismic model of the star
with oscillation frequencies that are as similar as possible to the observed frequencies.
This process is computationally expensive as a large
number of evolutionary tracks with different stellar parameters
need to be computed.
A seismic model provides oscillation \emph{kernels} that
allow the application of inversion techniques (see Sect. \ref{invmet}) to
determine the approximate rotation rates in different regions of the
star. This is the approach taken by \citet{deheuvels2012,deheuvels2014},
\citet{triana2015}, and \citet{dimauro2016}.

\citet{goupil2013} developed a powerful method that allows estimating the rotation rates of both core and envelope, without recurring to any seismic model,
by considering the relative amounts of `trapping' of a set of mixed modes, which as they showed are 
linearly related to the rotational splittings $\delta$
(see Section \ref{goupil}). The method relies solely on observed quantities  as inputs and particularly, on an estimate
of the asymptotic period spacing $\Delta\Pi_1$ that pure high-order g modes would have in the asymptotic
regime \citep{mosser2012b}.

This latter approach
was taken by \citet{deheuvels2015} in their sample of seven core He-burning red giants.
In that study, a seismic model was obtained for one of the stars in the
sample (KIC\,7581399) and was used to compute rotation rates through inversions.
Then the authors adapted the method of \citet{goupil2013}, and the rotation rates 
thus obtained were in 
very good agreement with the inversions based on the seismic model. After
assessing the validity of the Goupil approach in this way, no further seismic
modeling was attempted for the other targets, and the
corresponding rotation rates reported
come solely from the use of the adapted method.

\citet{mosser2015} provided additional insight into the relationship between the
relative amount of trapping of a mode (quantified by the parameter $\zeta$) and the
observed period spacing $\Delta P$ between consecutive mixed modes. The modification 
of Goupil's formula introduced by \citet{deheuvels2015} is 
exactly the same expression as was found by \citet{mosser2015} to represent the ratio
$\Delta P/\Delta\Pi_1$.

{ The model-independent method used by \citet{goupil2013} has been compared with inversions based
on seismic models for only two targets: the core helium-burning red giant KIC7581399 \citep{deheuvels2015},
  and the early red giant KIC4448777 \citep{dimauro2016}}. Although a wider comparison
of the method using targets in different evolutionary stages is desirable, we offer
comparisons of the { model-independent} method against inversions for our 13 targets, which share similar
evolutionary stages { with KIC4448777}. Our targets
are a subset of the original selection of 19 low-mass, low-luminosity 
red giants studied by \citet{corsaro2015} using
the Bayesian inference technique \textsc{D\large{iamonds}}
 \citep{corsaro2014} to detect and identify pulsation frequencies. 
\citet{hernandez2016} performed a grid-based search for models designed specifically 
to constrain the age, mass, and initial helium content of all 19 targets. 
In the present work we search for optimal seismic models for the 13 targets
that exhibit rotational splittings (Section \ref{seismic_mods}) and
employ Bayesian inference and inversion techniques 
(Section \ref{invmet}) to obtain average rotation rates. We compare these results 
with those obtained by { the model-independent} method as 
implemented by \citet{deheuvels2015} and by \citet{mosser2015} using their expressions
for the trapping parameter $\zeta$ (Section \ref{inrot}).
{Additionally, we use the idea proposed recently by \citet{klion2016} that provides
a way to localize the differential rotation of a red giant (whether in the radiative core or in the
convective envelope of the star), provided that the rotation rate of the envelope
is known by other means.}


\section{Rotation rate averages using the trapping parameter $\zeta$}
\label{goupil}

The parameter $\zeta$ gives an indication of how strongly a given stellar
 pulsation mode is localized, or`trapped',
inside the radiative core. It is defined as the ratio of the mode
inertia computed within the {\it g} -mode cavity $I_g$ \citep[][]{goupil2013} 
to the total mode inertia $I$:

\begin{linenomath*}
\begin{equation}
\zeta \equiv \frac{I_g}{I} = 
\frac{ \int_{r_1}^{r_2} \rho r^2 \left[\xi_r^2 +l(l+1) \xi_h^2 \right]\,\mathrm{d}r }
{ \int_0^{R_*} \rho r^2 \left[\xi_r^2 +l(l+1) \xi_h^2 \right]\,\mathrm{d}r }.
\label{zeta}
\end{equation}
\end{linenomath*}

\begin{figure*}
\begin{center}$
\begin{array}{cc}
\includegraphics[width=0.45\linewidth]{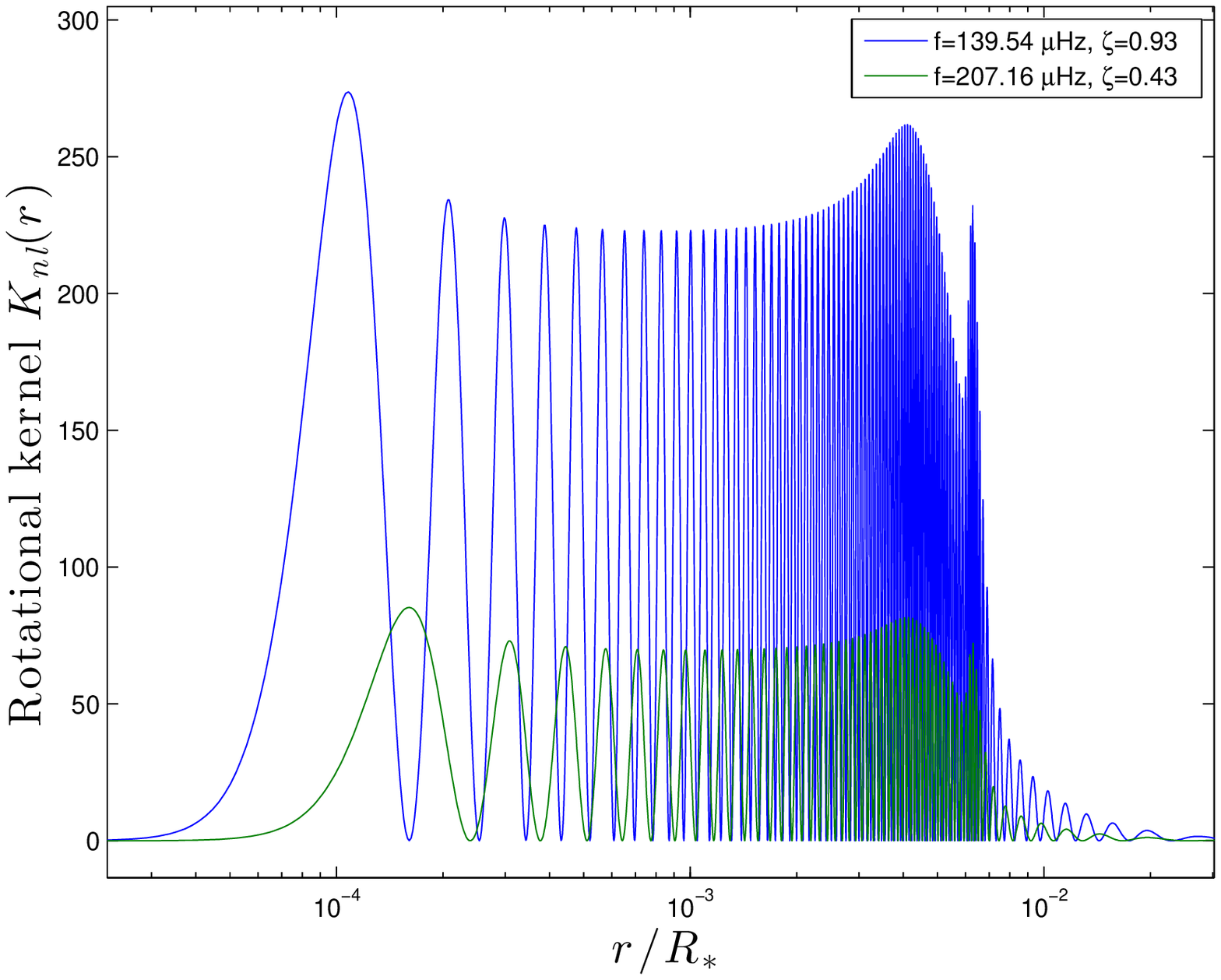}&
\includegraphics[width=0.45\linewidth]{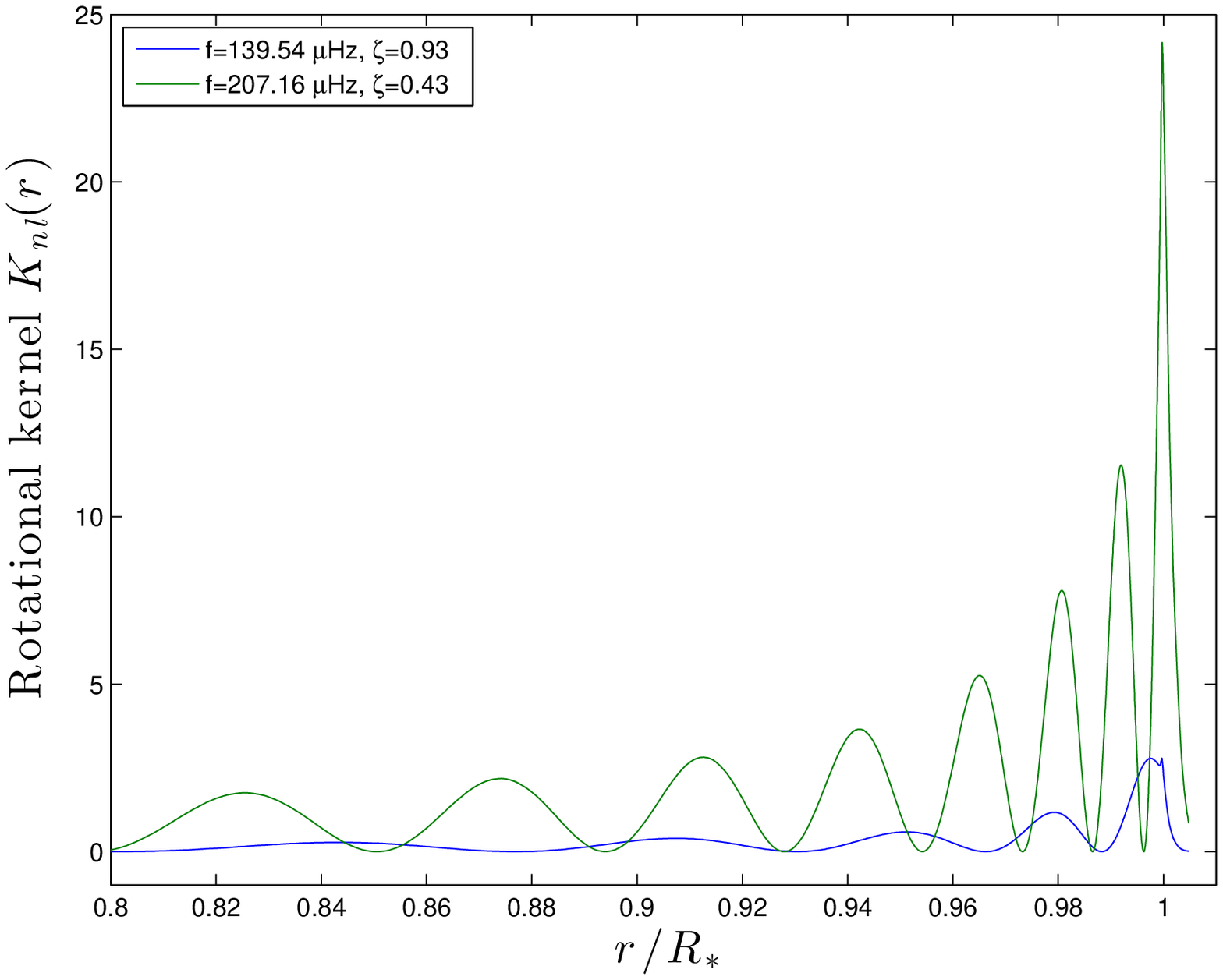}\\
\end{array}$
\end{center}
\caption{Rotational kernels of two dipole modes with
different values of the trapping parameter $\zeta$ 
(from the best model of KIC007619745). The left panel shows
the central region, and the right panel shows the region near the surface.}
\label{rotker}
\end{figure*}

In the expression above, $\rho$ is the density, $r_1$ and $r_2$ are the turning points of the {\it g} -mode
cavity, $l$ is the angular wave number of the mode, $\xi_r$ and $\xi_h$ are the vertical
and horizontal material displacement eigenfunctions, respectively, and $R_*$ 
is the stellar radius \citep{Aerts2010asteroseismology}.

Figure\,\ref{rotker} shows the rotational kernels $K_{nl}$ of two mixed modes with
different $\zeta$ using a seismic model for KIC007619745. 
Modes with $\zeta$ close to one are gravity-dominated mixed modes,
while modes with $\zeta$ close to 0.5 correspond to pressure-dominated mixed modes.

The rotational splittings $\delta_i$ are linearly related to $\zeta_i$ { for each mode $i,$}
as shown by \citet{goupil2013}, which we also refer to for further details. The coefficients of this linear
relationship are related directly to the rotational averages { across the core and the envelope through}  
\begin{linenomath*}
\begin{equation}
\delta=\left(\frac{\Omega_g}{2}-\Omega_p\right)\zeta+\Omega_p,
\label{goupil_eq}
\end{equation}
\end{linenomath*}
where $\Omega_p$ represents the average rotation rate
in the envelope (approximated by the {\it p} -mode cavity), and $\Omega_g$
represents the average rotation rate in the radiative core (approximated by
the {\it g} -mode cavity). { Following \citet{goupil2013},} they are defined as
\begin{linenomath*}
\begin{equation}
\label{avtur1}
\Omega_g=\frac{\int_0^{r_2}K(r)\,\Omega(r)\,\mathrm{d} r}{\int_0^{r_2}K(r)\,\mathrm{d}r}
\end{equation}
\end{linenomath*}
for the core, and
\begin{linenomath*}
\begin{equation}
\label{avtur2}
\Omega_p=\frac{\int_{r_2}^{R_*} K(r)\,\Omega(r)\,\mathrm{d} r}{\int_{r_2}^{R_*} K(r)\,\mathrm{d}r}
\end{equation}
\end{linenomath*}
for the envelope, where $r_2$ is the outer turning point in the g resonant cavity{ and $K(r)$ is the rotational kernel
of a mixed-mode. Our tests with the rotational profiles considered in Section \ref{tests} show that while the core averages are essentially independent of the particular mixed-mode chosen, the envelope averages
differ appreciably across modes with $\zeta\gtrsim 0.9$ (i.e., gravity-dominated modes). Using kernels from mixed modes
 with $\zeta\lesssim 0.85$ results in envelope averages with minimal variability.}

\subsection{Estimation of the trapping parameter $\zeta$}
\label{est_zeta}

The expression for $\zeta$ given by Eq.\,(\ref{zeta}) { in principle
requires} knowing the material displacement
eigenfunctions $\xi_{r,h}(r)$ for each mode, which are only available after the computationally
expensive process of deriving a seismic model of the star.
However, \citet{goupil2013} used an asymptotic {analysis} method based on the work of \citet{shibahashi1979}
 to show that $\zeta$ can
in principle be estimated using observational data alone. The expression for $\zeta$ was later refined by
\citet{deheuvels2015} and \citet{mosser2015}. In what follows, we briefly recall the method and the main formulae,
we refer to the original works for further details.

\begin{figure}
\centerline{
\includegraphics[width=1.0\linewidth]{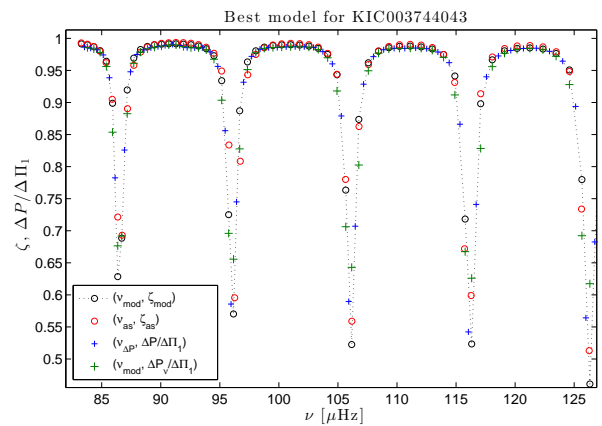}
}
\caption{ Trapping parameter as described by the true model, $\zeta_\mathrm{mod}$ (Eq.~\ref{zeta}) as a function
of the true frequencies $\nu_\mathrm{mod}$ (dashed black line and black circles). Red circles
represent the trapping parameter $\zeta_\mathrm{as}$ as approximated by 
Eqs.~\ref{zeta_eq} and~\ref{unno3} and using $\nu_\mathrm{as}$ for the mode frequencies as determined
by Eq.~\ref{unno1}. The scaled period spacing $\Delta P/\Delta\Pi_1$ between any two consecutive true 
frequencies $\nu_{\mathrm{mod},1}$ and $\nu_{\mathrm{mod},2}$ is plotted at the abscissa
$\nu_{\Delta P}=2/(\nu^{-1}_{\mathrm{mod},1}+\nu^{-1}_{\mathrm{mod},2})$ (blue crosses). $\Delta P_\nu$ represent
the interpolated value of the two adjacent period spacings around each frequency $\nu_\mathrm{mod}$ (green crosses).}
\label{zeta_comp}
\end{figure}

The method consists of finding aproximate JWKB solutions
for the material displacement eigenfunctions $\xi_{r,h}(r)$ in the two separate {\it p} and
{\it g} cavities of the star. { Matching
the solutions in the evanescent zone requires \citep{unno1989}
\begin{linenomath*}
\begin{equation}
\tan \theta_p=q\,\tan\theta_g,
\label{unno1}
\end{equation}
\end{linenomath*}
where $q$ is the coupling constant between the \textit{p}- and  \textit{g}-mode cavities, and the phases $\theta_{p,g}$ are defined through
\begin{linenomath*}
\begin{equation}
\theta_g=\int_{r_1}^{r_2}k_r\,\mathrm{d}r,\,\,\theta_p=\int_{r_3}^{r_4}k_r\,\mathrm{d}r,
\label{unno2}
\end{equation}
\end{linenomath*}
where $r_3$ and $r_4$ are the inner and outer turning points of the \textit{p}-mode cavity.
According to \citet{mosser2012b}, asymptotic analysis yields
\begin{linenomath*}
\begin{equation}
\theta_p=\frac{\pi}{\Delta\nu}(\nu-\nu_p),\,\,\theta_g=\pi\left( \frac{1}{\Delta\Pi_1\,\nu}-\epsilon_g \right),
\label{unno3}
\end{equation}
\end{linenomath*}
where $\nu_p$ is the frequency of the theoretical $l=1$ pure \textit{p} modes, which are related to the radial ($l=0$) modes $\nu_{n,0}$ through
 \begin{linenomath*}
\begin{equation}
\nu_p=\nu_{n,0}+\left(\frac{1}{2}-d_{01}\right)\Delta\nu.
\label{unno4}
\end{equation}
\end{linenomath*}
In turn, the radial modes can be expressed as a function of the radial order $n$ involving the parameters $\epsilon_p$, $\alpha$,
and the large frequency separation $\Delta\nu$ as follows \citep{mosser2013}: 
\begin{linenomath*}
\begin{equation}
\nu_{n,0}=\left[n+\epsilon_p+\frac{\alpha}{2}(n-n_\mathrm{max})^2\right]\Delta\nu,
\label{unno5}
\end{equation}
\end{linenomath*}
where $n_\mathrm{max}\equiv \nu_\mathrm{max}/\Delta\nu-\epsilon_p$. The approximate
 expression for the trapping parameter, denoted here as $\zeta_\mathrm{as}$ , reads
\begin{linenomath*}
\begin{equation}
\zeta_\mathrm{as}=
\left[ 1 + \frac{\nu^2\,\Delta\Pi_1}{q\,\Delta\nu}\frac{\cos^2\theta_g}{\cos^2\theta_p} \right]^{-1}.
\label{zeta_eq}
\end{equation}
\end{linenomath*}
A total of seven parameters are required here: the coupling constant $q$, 
the offsets $\epsilon_{p,g}$, the mean large frequency
separation $\Delta\nu$, the asymptotic period spacing $\Delta\Pi_1$, $\alpha,$ and
$d_{01}$. In practice, the optimal parameters $\epsilon_p$, $\alpha$, and $\Delta\nu$ are determined first
by fitting the observed radial modes to Eq.~\ref{unno5}. Then, the optimal parameters $q$, $\Delta\Pi_1$, $\epsilon_g$,
and $d_{01}$ that best reproduce the observed $l=1$ mode frequencies can be found 
by a downhill simplex method. This requires solving Eq.~\ref{unno1} for the mode frequencies $\nu$ at each search step
with a particular $(q,\,\Delta\Pi_1,\,\epsilon_g,\,d_{01})$ combination, using a Newton method to find the
roots of the equation, for example.}

The observed rotational splittings $\delta$ are expected to be linearly related to $\zeta_\mathrm{as}$ 
, and a linear fit of $\delta$ as a function of $\zeta_\mathrm{as}$ leads to an estimate of the average envelope rotation and the envelope core rotation
through Eq.\,\ref{goupil_eq}.

\citet{mosser2015} obtained a result in their search of an expression
for the mixed-mode relative period spacings 
$\Delta P/\Delta \Pi_1$  that exactly matched the expression
for $\zeta_\mathrm{as}$ (Eq.~\ref{zeta_eq}) found by \citet{deheuvels2015}. Thus, we have
\begin{linenomath*}
\begin{equation}
\zeta_\mathrm{as}=\frac{\Delta P}{\Delta \Pi_1}.
\label{gdm}
\end{equation}
\end{linenomath*}
The above is a reflection of the fact that the rotational
splittings follow the same distribution as the period spacing, because both are
determined by the coupling between pressure and gravity terms.
As a bonus, Equation\,\ref{gdm} provides a simple and direct way to estimate 
$\zeta$ from observations{ without the need of the optimal parameters mentioned above
(which can be slightly time consuming computationally) 
with the exception of $\Delta\Pi_1$. Some care must be taken because the period spacing
between two consecutive mixed dipole modes $\Delta P(n,n+1)=\nu^{-1}_{n+1}-\nu^{-1}_n$ is defined properly at 
$\nu=2/(\nu^{-1}_{n+1}+\nu^{-1}_n)\equiv\nu_{\Delta P}$. In order to assign a 
$\Delta P$ to each mode $\nu_n$, we therefore interpolate the two adjacent period spacings
 $\Delta P(n,n+1)$ and $\Delta P(n-1,n) $ linearly.  Similarly, when performing
  linear fits of $\delta$ vs. $\zeta_\mathrm{as}$, it is advisable to also include
the interpolated rotational splitting at each location $\nu_{\Delta P}$ using the two correspondingly adjacent
values of $\delta$ to minimize biases, see Fig.~\ref{zeta_sp_fig}.

In Fig.~\ref{zeta_comp} we show that the trapping parameter $\zeta_\mathrm{mod}$ as derived from a known seismic model is indeed well
approximated by either $\zeta_\mathrm{as}$ derived using asymptotic analysis, Eq.~\ref{zeta_eq}, or by the simpler
expression given by Eq.~\ref{gdm}.}

Assuming
that the errors on the frequencies and the splittings are normally
distributed, we can sample randomly from them and proceed to
compute{ interpolated} splittings and spacings as explained earlier. A linear fit to these points leads 
to estimates of $\Omega_g$ and $\Omega_p$ according to Eq.~\ref{goupil_eq}. 
By repeating these steps many times, we can obtain the 
distributions associated with $\Omega_g$ and 
$\Omega_p$ { and their associated errors}.

\subsection{Bayesian inference}

If a seismic model providing the oscillation eigenfunctions $\xi_{r,h}$ is available,
 the trapping parameter can be computed from Eq.\,\ref{zeta}, which we denote now as $\zeta_\mathrm{mod}$.
With two sets of inputs, that is, the splittings $\delta_i$ and the trapping parameters $\zeta_{\mathrm{mod},i}$,
we can set out to perform a Bayesian fit using Eq.\,\ref{goupil_eq} as the model. To accomplish this,
 we first compute a Gaussian log-likelihood function defined as \citep[see also][]{corsaro2013}
\begin{linenomath*}
\begin{equation}
\Lambda( \Omega_g,\,\Omega_p) = \Lambda_0 - \frac{1}{2} \sum^{N}_{i=1} \left[ \frac{ \Delta_i(\Omega_g\,,
 \Omega_p) }{\epsilon_i} \right]^2 \, ,
\end{equation}
\end{linenomath*}
where $N$ is the total number of rotational splittings $\delta_i$ from the observations
 (one for each m-multiplet of mixed modes),
  $\Delta_i(\Omega_g,\,\Omega_p)$ 
  are the residuals given as the difference between the observed and the modeled splittings,
   $\epsilon_i$ the corresponding uncertainty, and
\begin{equation}
\Lambda_0 = - \sum^{N}_{i=1} \ln {\sqrt{2 \pi}\, \epsilon_i} \, ,
\end{equation}
a constant term. We multiply the likelihood distribution by the prior distributions (uniform or flat in this case,
  in the range $\Omega_g \in \left[ 2, 8 \right]\,\mu$Hz, and $\Omega_p \in \left[ 0, 3 \right]\,\mu$Hz),   obtaining a posterior probability density distribution. 
   By marginalizing the bidimensional posterior into two one-dimensional probability density distributions,
    we obtain estimates for $\Omega_p$ and $\Omega_g$ that are the medians of the two one-dimensional distributions.
     The corresponding error bars are the Bayesian credible intervals computed as explained in \citet{corsaro2013}. 
      This statistical approach may provide similar results to a least-squares fit, but it
      is conceptually very different. One of the main differences is that it is able to 
      incorporate any a priori knowledge on the estimated parameters that we may have, for instance, in the form
      of the prior distributions described above.

\begin{figure}[h!]
\centerline{
\includegraphics[width=1.0\linewidth]{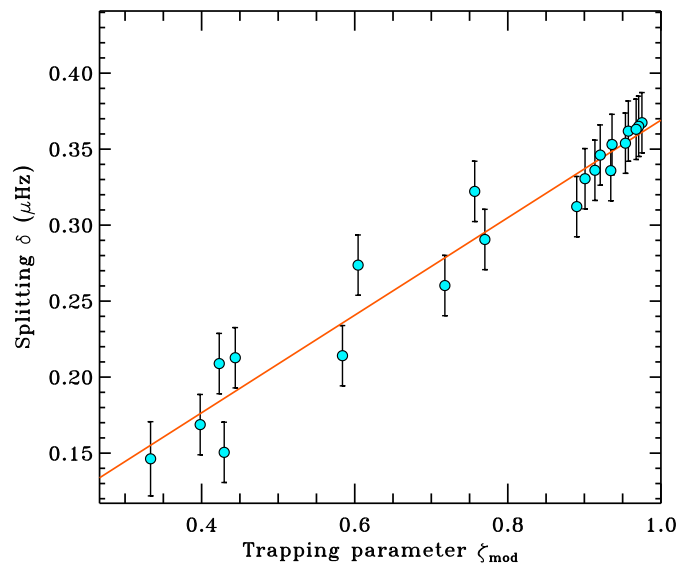}
}
\caption{Illustrative use of a Bayesian fit to compute rotation rates for KIC007619745. 
The trapping parameters $\zeta_\mathrm{mod}$ are obtained from the eigenmodes as computed by GYRE.
}
\label{ec}
\end{figure}

An example of this Bayesian fit is shown in Figure~\ref{ec} for the star KIC007619745 (solid orange line), with 1$\sigma$ error
 bars overlaid. The results from this method for all targets are included in Table 1 under the `Bayes' heading.

\section{Inversion methods}
\label{invmet}
In this section we present the methods we used to obtain rotation rate
averages, which are all based on the oscillation kernels provided by the seismic models
described in Section \ref{seismic_mods}.{ Our treatment is based on \citet{Aerts2010asteroseismology}, who offered an extended presentation of the methods discussed below.}

The so-called \emph{forward} problem states that the rotational splittings
$\delta_{nlm}$ of an oscillation mode with radial, angular, and azimuthal wavenumbers
$n,l,m$ can be computed through Eq.\,(\ref{forward}). Explicitely, the kernels are computed from
the material displacement eigenfunctions via 
\begin{linenomath*}
\begin{equation}
K_{nl}=\frac{1}{I}\left[ \xi_r^2+l(l+1)\,\xi_h^2 -2\xi_r\xi_h-\xi_h^2\right]r^2\rho,
\end{equation}
\end{linenomath*}
where $I$ is the mode inertia (see Eq.\,\ref{zeta}). The constant 
$\beta_{nl}$ is given by
\begin{linenomath*}
\begin{equation}
\beta_{nl}=\frac{1}{I} \int_0^{R_*}\left[ \xi_r^2+l(l+1)\xi_h^2 -2\xi_r\xi_h-\xi_h^2\right]r^2\rho\, \mathrm{d}r.
\end{equation}
\end{linenomath*}

The {\it inverse} problem consists of determining the unknown inversion
coefficients $c_i(r)$ satisfying
\begin{linenomath*}
\begin{equation}
\label{c_i}
\bar \Omega(r)=\sum_{i=1}^M c_i(r) \frac{\delta_i}{m\,\beta_i},
\end{equation}
\end{linenomath*}
where $\bar\Omega(r)$ is the \emph{predicted} internal rotation rate of the star,
$M$ is the number of observed splittings, and $i$ denotes the collective indices $(n,l,m)$. { Clearly, the
inversion coefficients $c_i$ are not determined by Eq.~\ref{c_i}, which just states
 a linear relationship between the observed splittings and the predicted rotation profile. The
 $c_i(r)$ are determined by minimizing the difference between observed and predicted splittings, by
  minimizing of the resulting uncertainties, or by adjusting the shape of the averaging kernels, as discussed below.}

The approximate rotational profile $\bar
\Omega(r)$ can be expressed in terms of the true profile $\Omega(r)$ by means of the 
\emph{averaging} kernels $\mathcal{K}(r',r)$, which are related to the kernels
$K_i(r)$ through $ \mathcal{K}(r',r)=\sum_{i=1}^M c_i(r') K_i(r)$ and fulfill
\begin{linenomath*}
\begin{equation}
\label{eq:avker}
\bar \Omega(r')=\int_0^{R_*} \mathcal{K}(r',r)\, \Omega(r)\, \mathrm{d}r.
\end{equation}
\end{linenomath*}
The averaging kernels $\mathcal{K}(r',r)$ should be localized around $r'$ as much
as possible, ideally resembling a delta function $\delta(r',r)$.
It is usually assumed that the observational errors $\epsilon_i$ are uncorrelated 
(as in, e.g., \citet{deheuvels2012} or \citet{dimauro2016}), so that
the variance of the predicted rotation rates can be estimated as
\begin{linenomath*}
\begin{equation}
\label{eq:variance}
\sigma^2\left[\bar\Omega(r)\right]=\sum_{i=1}^M c_i^2(r)\,\left(\frac{\epsilon_i}{\beta_i}\right)^2.
\end{equation}
\end{linenomath*}
The expression above accounts for the errors originating from the observations alone; it
 does not account for the inherent errors of the inversion process itself.

\subsection{Two-zone inversion models}

\begin{figure}
\begin{center}
\includegraphics[width=1.0\linewidth]{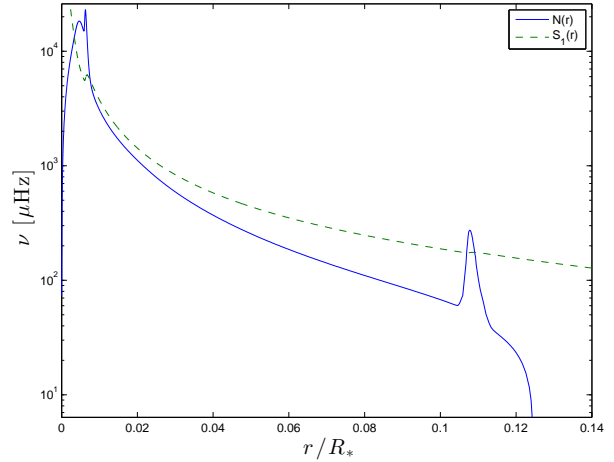}
\caption{Propagation diagram of the best model of KIC\,007619745. The
Brunt-V\"ais\"al\"a frequency $N(r)$ is plotted as a solid blue curve,
and the $l=1$ Lamb frequency $S_1(r)$ is plotted as a dashed green curve.
\textit{p}-modes with frequency $\nu$ are evanescent in the region where $\nu<S_1(r)$. Similarly,
\textit{g}-modes are evanescent wherever $\nu>N(r)$. The core-envelope boundary in our two-zone models
is located at $r_c\approx 0.1 R_*$ , which usually lies in the evanescent zone common
to both \textit{p}- and \textit{g}-modes.    
The peak of $N(r)$ near $r/R_*=0.11$ is associated with a chemically inhomogeneous layer in the radiative zone,
 left behind by convection as the boundary moves upward.}
\label{brunt}
\end{center}
\end{figure}

To obtain approximate averages of the core and envelope
rotation rates, we can make use of simple two-zone models where we assume an inner
zone extending from the stellar center to $r/R_*=x_c$ and an outer zone extending
from $r/R_*=x_c$ all the way to stellar surface at $r/R_*=1$, both zones
rotating uniformly with rates $\Omega_g$ and $\Omega_p$ , respectively. Our 13 targets
happen to be approximately at the same evolutionary stage, and therefore it is not surprising
that our seismic models show all their evanescent zones located
 at approximately the same radial locations (scaled by stellar radius)  
{We have chosen $x_c$ to coincide with the base of the convection zone for each target}, 
see Figure~\ref{brunt}.
We can determine the inversion coefficients $c_i$ associated with each
zone by finding the optimal $\Omega_g$ and $\Omega_p$ that minimize
\begin{linenomath*}
\begin{equation}
\label{chi2}
\chi^2=\sum_{i=1}^M \left(\frac{\bar \delta_i- \delta_i}{\epsilon_i}\right)^2,
\end{equation}
\end{linenomath*}
where $\epsilon_i$ are the observation errors and $\bar\delta_i$ are the predicted
splittings associated with the { two-zone} rotation profile composed of $\Omega_g$ and $\Omega_p$. 
{ The averages are determined by enforcing $\partial(\chi^2)/\partial\Omega_{g,p}=0$ after substituting
$\chi^2$ using Eqs.~\ref{forward} and \ref{chi2}.} 
Results from this method, with kernels from the seismic models discussed below, are presented on
Table~1 under the `Two-zone' heading.

\subsection{Subtractive optimally localized averaging}
One of the differences between
the subtractive optimally localized averaging (SOLA) method \citep{pijpers1994}
and the method described above is that with SOLA we do not minimize $\chi^2$ , but instead,
the method chooses the optimal linear combination of the
 inversion coefficients $c_i$ such
that the averaging kernels $\mathcal{K}(r',r)$ resemble a given
\emph{target} function
$T(r',r)$ as closely as possible while keeping the variance
$\sigma^2(\bar \Omega)$ low. Thus, we minimize
\begin{linenomath*}
\begin{equation}
\label{eq:sola1}
\int_0^{R_*} \left[ \mathcal{K}(r',r)-T(r',r) \right]^2 \mathrm{d}r + \mu \sum_{i=1}^M c_i^2(r')\, \epsilon_i^2
\end{equation}
\end{linenomath*} 
at each $r'$, with the additional constraint $\int_0^{R_*} \mathcal{K}(r',r)\,
\mathrm{d}r=1$. The target function that we have chosen in this study is a Gaussian with unit norm, 
centered on $r=r'$ with adjustable width $s$:
\begin{linenomath*}
\begin{equation}
T(r',r)=N\,\mathrm{e}^{-{\left(\frac{r'-r}{s}\right)}^2},
\end{equation}
\end{linenomath*}
$N$ being a normalization factor.
In addition to the free parameter $\mu$ in Eq.\,\ref{eq:sola1}, we can also adjust the
shape of the target function $T$ by adjusting the width $s$. The problem reduces to solving the linear set
of $M$ equations ($i=1,\ldots,M$) for each radial location $r'$:
\begin{linenomath*}
\begin{equation}
\label{eq:sola2}
\sum_{k=1}^M W_{ik}\, c_k(r')=\int_0^{R_*} K_i(r)\, T(r',r)\, \mathrm{d}r,
\end{equation}
\end{linenomath*}
where $W_{ik}=\int_0^{R_*} K_i(r)\, K_k(r)\, \mathrm{d}r + \mu\, \delta_{ik}\,
\epsilon_i^2$, together with the constraint $\sum_k c_k(r')=1,$ { which is implemented via
Lagrange multipliers. Given a set of kernels $K_i(r)$ and the two parameters $(s,~\mu)$, the inversion coefficients
$c_i(r')$ are determined by solving the set of $M$ equations given by Eq.~\ref{eq:sola2}. We note that the observed splittings are
not involved in determining the $c_i(r')$, they determine the predicted rotation rate $\bar\Omega(r')$ via Eq.~\ref{c_i}. }

\section{Testing the methods}
\label{tests}

Before applying the methods described above to determine rotation rates for the stars in our sample, it is
desirable to have an idea of how the methods perform under controlled situations. We choose a specific seismic
model (of KIC007619745, obtained as explained in the next section) as the `true' model.
 Then, we consider six different rotation profiles
and compute the exact rotational splittings in each case via Eq.~\ref{forward}. We also compute the `true'
rotational averages for both the core and the envelope for each case using Eqs.~\ref{avtur1} and \ref{avtur2}.

For the first set of rotation profiles we adopted the functional form used by \citet{klion2016}:
we assume that the inner region of the star rotates uniformly with a rate $\Omega_c$ from the
center and up to 1.5 times $r_H$, the outer radius of the hydrogen burning shell. Then, from 1.5 $r_H$ and up
to the base of the convective zone ($r_\mathrm{rcb}$), the star follows a uniform rotation rate $\Omega_m$. 
The remainder of the star
rotates according a power-law profile. Thus\begin{linenomath*}
\begin{equation}
\label{klionprof}
\Omega(r)=
\begin{cases} 
      \Omega_c & r \leq 1.5\, r_H,\\
      \Omega_m & 1.5\,r_H < r \leq r_\mathrm{rcb},\\
      \Omega_e\left(\frac{R_*}{r}\right)^\alpha & r > r_\mathrm{rcb},
\end{cases}
\end{equation}
\end{linenomath*}
where 
\begin{linenomath*}
\begin{equation}
\alpha=\frac{\log(\Omega_m/\Omega_e)}{\log(R_*/r_\mathrm{rcb})}.
\end{equation}
\end{linenomath*}
 The exponent $\alpha$ is so chosen to ensure the continuity of $\Omega(r)$ at $r=r_\mathrm{rcb})$. 
This functional form is useful to adjust the location of the differential rotation.
  By setting $\Omega_m=\Omega_e$ all the differential rotation in the star is localized
 at $r_\mathrm{rcb}$, inside the radiative region, and if $\Omega_m=\Omega_c,$ the differential rotation is all contained in the
 convective envelope. We have kept $\Omega_c$ and $\Omega_e$ fixed at $0.7\,\mu$Hz and $0.1\,\mu$Hz, respectively. We consider
 three different values for $\Omega_m$: $0.7\,\mu$Hz, $0.4\,\mu$Hz, and $0.1\,\mu$Hz.
 
\begin{figure}
\centerline{
\includegraphics[width=1.0\linewidth]{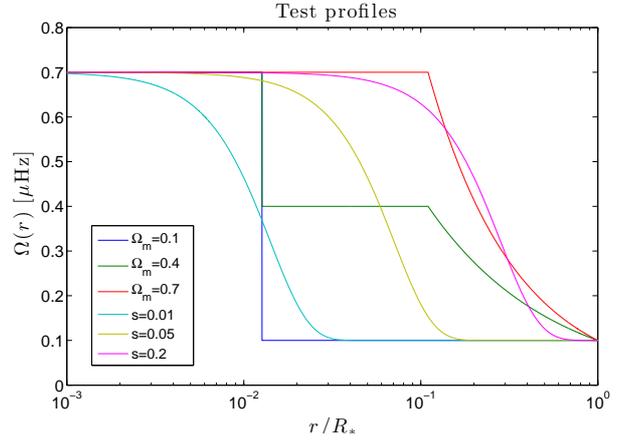}
}
\caption{ Synthetic rotation rate profiles used for testing. $\Omega_m$ refers to the rotation rate between 1.5 $r_H$ and the base
convective zone, see Eq.~\ref{klionprof}. $s$ refers to the width of a Gaussian profile as defined in Eq.~\ref{gaussprof}.}
\label{testprofs}
\end{figure}

 The other set of rotation profiles are Gaussians of different widths $s$ plus a constant term $B$:
 \begin{linenomath*}
\begin{equation}
\label{gaussprof}
\Omega(r)=A\,\mathrm{e}^{-\frac{1}{2}\left(\frac{r/R_*}{s}\right)^2}+B,
\end{equation}
\end{linenomath*}
we set $A=0.6\,\mu$Hz, $B=0.1\,\mu$Hz and three different widths $s$: 0.01, 0.05 and 0.2. See Fig.~\ref{testprofs}.

\begin{figure}[h]
\label{zeta_sp_fig}
\centerline{
\includegraphics[width=1.0\linewidth]{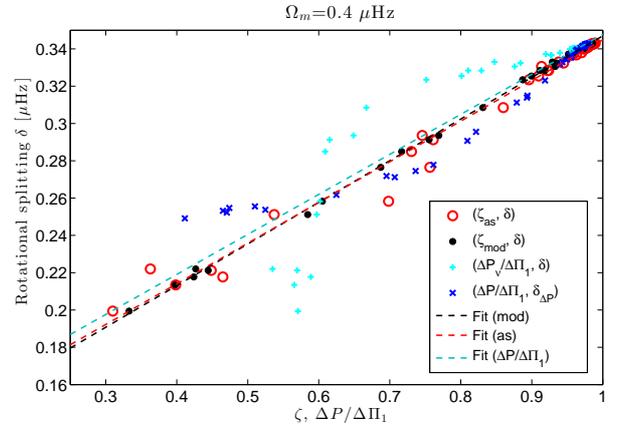}
}
\caption{ Rotational splittings using the best seismic model of KIC007619745 
for one of the test profiles as defined in Eq.~\ref{klionprof} with
 $\Omega_c=0.7~\mu$Hz, $\Omega_m=0.4~\mu$Hz, and $\Omega_e=0.1~\mu$Hz. The splittings $\delta_{\Delta P}$ (dark blue crosses)
 are the interpolated splittings as explained in Section \ref{est_zeta}. Dotted lines are linear fits providing estimates of the rotation
 rate averages in the \textit{g} and \textit{p} cavities following Eq.~\ref{goupil_eq}.}
\label{zeta_sp}
\end{figure}

We compute the rotational splittings of the six test profiles using Eq.~\ref{forward} for {\it all} 
the $l=1$ mode frequencies of the true model within
$\pm3\,\Delta\nu$ from the frequency at maximum power $\nu_\mathrm{max}$ . Then we compute the optimal combination
of parameters $(q,~\Delta\Pi_1,~\epsilon_g,~d_{01})$ that best reproduces these model frequencies in order to
estimate $\zeta_\mathrm{as}$. As explained in Section~\ref{est_zeta}, we compute interpolated splittings
$\delta_{\Delta P}$ to correspond with each period spacing $\Delta P$ as well as interpolated period spacings
$\Delta P_\nu$ to correspond with each splitting $\delta$, see Fig.~\ref{zeta_sp}.

\begin{figure*}[h!]

\centerline{
\includegraphics[width=1.0\linewidth]{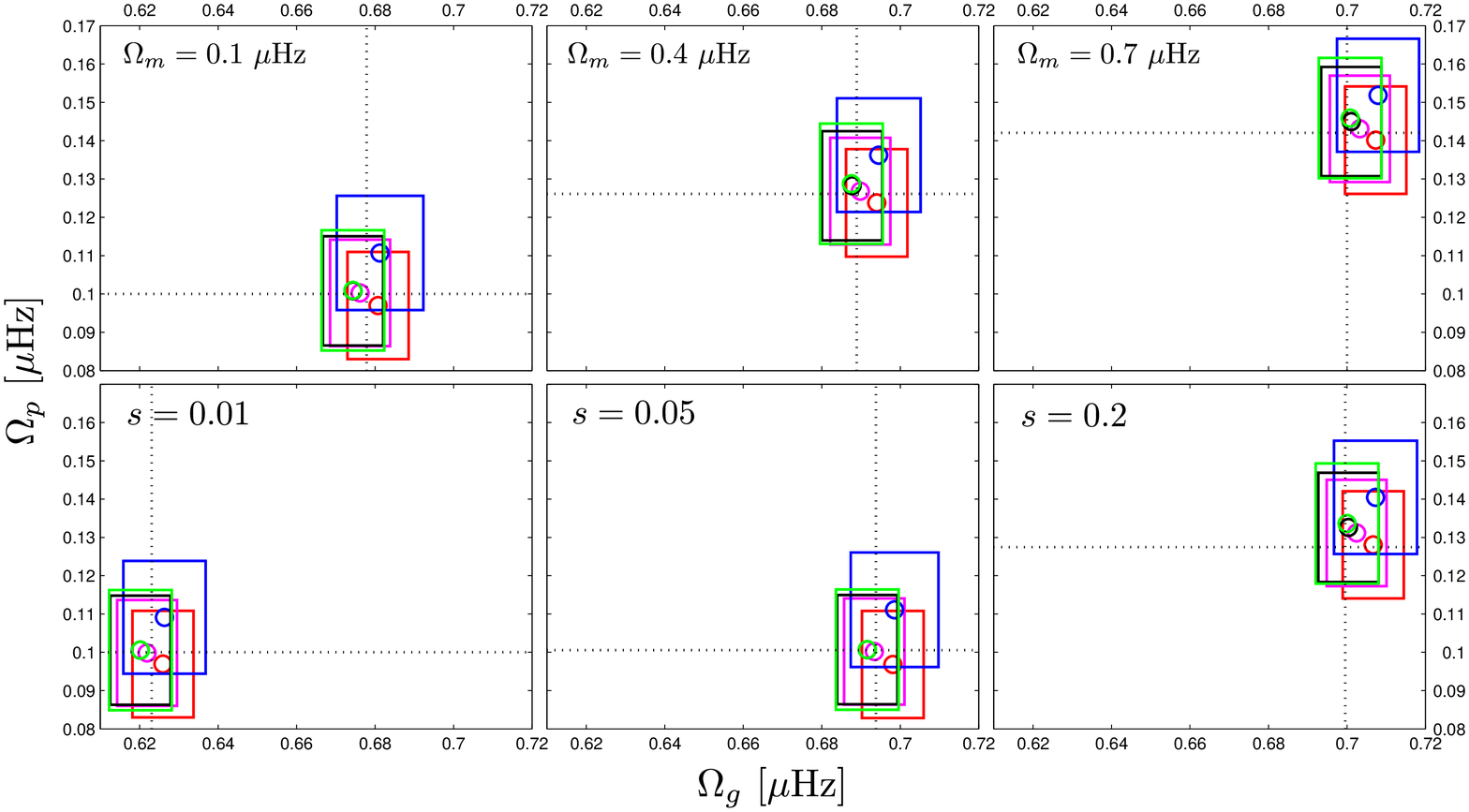}
}
\caption{ Comparison of the predicted rotation rate averages in the core ($\Omega_g$) and in the envelope ($\Omega_p$) using
the methods described in Sections~\ref{goupil} and \ref{invmet} based on the best seismic model
for KIC007619745. The top row corresponds to rotation profiles following
Eq.~\ref{klionprof}, and the bottom row corresponds to Gaussian rotation profiles according to Eq.~\ref{gaussprof}. The black
dashed lines mark the true averaged rotation rates as computed from Eqs.~\ref{avtur1} and \ref{avtur2}. Red represents
the rotation rates as determined from linear fits of ($\zeta_\mathrm{mod},\,\delta$), and magenta represents the rotation rates 
from linear fits of ($\zeta_\mathrm{as},\,\delta$). Blue corresponds to linear fits of $\delta$ and $\delta_{\Delta P}$ 
vs $\Delta P_\nu/\Delta\Pi_1$ and $\Delta P/\Delta\Pi_1$. Black are the predictions from two-zone inversions, and green are
the predictions from SOLA inversions. Results from Bayesian inference are not included here as they are essentially identical
to the two-zone models.}
\label{testnonoise}
\end{figure*}

Now we obtain the estimates of the rotation rates using each of the methods explained earlier.
The true averages are obtained following Eqs.~\ref{avtur1} and \ref{avtur2}. In the case of the SOLA
method, we computed the predicted rotation rates in two different locations: one at the surface of the
star ($r/R_*=1$), and the other well within the radiative core ($r/R_*=10^{-3}$). These SOLA-predicted
rotation rates are sensitive to the width $s$ of the target function. To obtain `calibrated'
values for $s,$ we therefore proceed first to compute the optimal two-zone model to determine the averages
$\Omega_g$ and $\Omega_p$. Then we compute the rotational splittings associated with this two-zone model and
use them as inputs for a SOLA inversion, adjusting the widths $s$ as necessary to make the
corresponding SOLA predictions at $r/R_*=10^{-3}$
and $r/R_*=1$ exactly match $\Omega_g$ and $\Omega_p$, respectively. With the widths $s$ determined in this way, we then proceed to compute
the SOLA-predicted averages using the splittings arising from the six test profiles.

We assume that the splittings all have the same uncorrelated Gaussian error distribution whose width matches
the mean $1\sigma$ error from the actual measurements. In the case of the inversions, the corresponding
uncertainties on the predicted rotation rates are computed via Eq.~\ref{eq:variance}. In the case of the
rotation rates estimated via linear fits involving $\zeta_\mathrm{mod,\,as}$ (see Fig.~\ref{zeta_sp}),
the resulting uncertainties are computed through a Monte Carlo simulation sampling
randomly and repeatedly for many times ($10^5$)
 from the normal distributions associated with each spliting $\delta$. For the fits
of $\delta,\,\delta_{\Delta P}$ vs $\Delta P_\nu/\Delta\Pi_1,\,\Delta P/\Delta\Pi_1$
we followed the same Monte Carlo approach, except that we also sampled randomly from
the distributions associated each mode frequency $\nu,$ which we set as having the
same standard deviation as the actual observed errors. We note that errors on the trapping parameter or systematic errors
incurred by the inversions are not considered.

The predicted rotation rates are shown in Fig.~\ref{testnonoise}. Although some scatter is
present, the predictions are in acceptable agreement with the true
averages across all methods. The test we just performed is an ideal situation,
the modes are properly identified, the rotational kernels are based on the true seismic model,
and the error distributions associated with the frequencies and splittings are centered exactly
on the true values. Any deviation from this ideal situation will bring additional scatter
to the predicted rotation averages.

The averaging kernels $\mathcal{K}(r)$ from the SOLA inversions at the two
radial locations mentioned above, although well resolved with respect to each other, are essentially identical
to the averaging kernels of the two-zone models at the corresponding zone, which suggests already that
no more than two reasonably well-defined averages can be obtained using SOLA inversions.

Recently, \citet{klion2016} proposed a method aiming to determine the region in a red giant star where differential rotation
is concentrated. They note that the minimum normalized splitting, $\mathrm{min}(\delta/\mathrm{max}(\delta))$ can be used to distinguish between 
rotation profiles with differential rotation localized in the
core from those with differential rotation localized mostly in the envelope. This is precisely the motivation
of the functional form of the rotation profiles given by Eq.~\ref{klionprof}. For a fixed ratio $\Omega_c/\Omega_e$ in these
profiles, the quantity $\mathrm{min}(\delta/\mathrm{max}(\delta))$ follows a one-to-one correspondence with $\Omega_m$ that determines
the location of the differential rotation either in the core or in the envelope.

\citet{dimauro2016} were able to resolve three rotation rates in three distinct radial locations of KIC4448777, two of them
 within the radiative core, which allowed them
to conclude that there is a steep gradient in the rotation there, thus localizing the differential
rotation of the star inside the radiative core. The target in that study is very similar to the targets
in our target selection, sharing similar evolutionary stages, therefore it may be possible in principle to resolve the rotation
rate in at least two points inside the core. Unfortunately, the set of rotational kernels for all of our targets is not suitable for obtaining more than an averaged value across the core (see Fig.~\ref{cumm}). The reason for this
is not evident a priori, and although it deserves special attention, it is beyond the scope of the present study.

\begin{figure}[h!]

\centerline{
\includegraphics[width=1.0\linewidth]{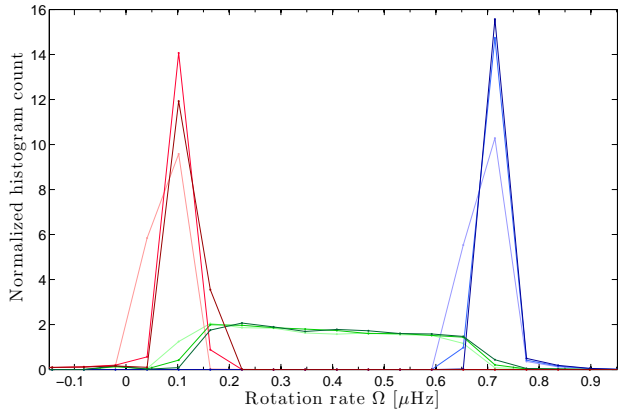}
}
\caption{ Histogram associated with the predictions for $\bar\Omega_{c,m,e}$ (blue, green, and red tones,
respectively).
The true (test) profile is described by Eq.~\ref{klionprof} with fixed $\Omega_c=0.7~\mu$Hz, $\Omega_e=0.1~\mu$Hz.
The light colors represent the histogram obtained when $\Omega_m=0.1~\mu$Hz, medium colors correspond to
$\Omega_m=0.4~\mu$Hz, and dark colors represent $\Omega_m=0.7~\mu$Hz.
Essentially identical distributions for $\bar\Omega_m$ result, regardless of the choice of $\Omega_m$.
While $\Omega_c$ and $\Omega_e$ are recovered properly, no good prediction for $\Omega_m$ can be made.}
\label{noresolv}
\end{figure}

We can use the idea by \citet{klion2016} to determine whether we can indeed localize the differential rotation. We consider a
rotation profile with some preestablished values for the set $(\Omega_c,\Omega_m,\Omega_e)$
 and compute its associated splittings. We consider splittings from {\it all} $l=1$ mixed modes within 
 $\pm3\,\Delta\nu$ of $\nu_\mathrm{max}$.
 With this set of splittings as input and
considering errors on them matching the actual observed errors, we proceed to find the optimum combination of
predicted 
$(\bar\Omega_c,\bar\Omega_m,\bar\Omega_e)$ that minimizes the difference between the input and the predicted splittings.
To minimize and obtain estimates of the three parameters, we use a downhill simplex method. Then
 we make use once again of a Monte Carlo simulation sampling randomly from the normal distributions associated with the
input splittings to obtain distributions for $(\bar\Omega_c,\bar\Omega_m,\bar\Omega_e)$. If enough
information is indeed contained in the set of input splittings to constrain $\bar\Omega_m$, then this should reflect in a
sharply peaked distribution for it. Figure \ref{noresolv} shows the results of this experiment. We set
$\Omega_c=0.7~\mu$Hz, $\Omega_e=0.1~\mu$Hz, and vary $\Omega_m\in\{0.1,~0.4,~0.7\}~\mu$Hz. The predicted
values for $\bar\Omega_c$ and $\bar\Omega_e$ are close to the true values, but the probability density 
distribution for $\bar\Omega_m$ is wide and practically flat, thus no reliable prediction for $\Omega_m$ is possible
and the differential rotation cannot be localized properly. Essentially identical distributions of $\bar\Omega_m$
are obtained regardless of the choice of $\Omega_m$. This is a consequence
of the magnitude of errors in the splittings together with the characteristics of the rotational kernels. Only if we artificially
reduce the errors by an order of magnitude or less, a reasonable value of $\bar \Omega_m$ can be recovered. 
All of the 13 targets in our sample exhibit this undesirable behavior.   

\begin{table*}[h!]
\caption{Selected stellar parameters from observations and best seismic models. For each target, the top row contains observed
quantities derived from {\em Kepler} data and bottom row contains quantities extracted from the best seismic model.}
\label{latabla}
\small     
\centering                          
\begin{tabular}{l D{.}{.}{1.7} l l l D{.}{.}{1.2} l l l l}        
\hline\hline                  
KIC id & \multicolumn{1}{c}{$\Delta\nu$ ($\mu$Hz)} & $\nu_\mathrm{max}$ ($\mu$Hz) & $\Delta\Pi_1$  (s) & $T_\mathrm{eff}$ (K) & \multicolumn{1}{c}{[Fe/H]$_\mathrm{ini}$} & $Y_\mathrm{ini}$ & $f_{\mathrm{ov}}$ ($H_P$) & $M/M_{\odot}$ & $R/R_{\odot}$ \\
\hline
\multirow{2}{*}{003744043} &  9.90\pm0.05 & $112.52\pm0.25$ & $76.0\pm1.3$ & $5112\pm172$ &       &       &        & $1.28\pm0.18$     &       \\
                           & 10.01          & 112.84          & 75.64        & 5014         & -0.22 & 0.300 & 0.0223 & 1.21              & 5.945 \\ \hline
\multirow{2}{*}{006144777} & 11.01\pm0.06 & $129.69\pm0.24$ & $79.3\pm1.4$ & $4746\pm186$ &       &       &        & $1.26\pm0.18$     &       \\
                           & 11.18          & 130.92          & 79.63        & 4820         & 0.02  & 0.270 & 0.0353 & 1.15              & 5.431 \\ \hline
\multirow{2}{*}{007060732} & 10.94\pm0.05 & $132.29\pm0.24$ & $72.8\pm1.4$ & $4790\pm164$ &       &       &        & $1.33\pm0.19$     &       \\
                           & 11.04          & 133.07          & 78.28        & 5056         & -0.22 & 0.285 & 0.0190 & 1.37              & 5.821 \\ \hline
\multirow{2}{*}{007619745} & 13.13\pm0.07 & $170.82\pm0.25$ & $79.2\pm1.3$ & $5126\pm169$ &       &       &        & $1.58\pm0.21$     &       \\
                           & 13.28          & 169.58          & 79.71        & 4969         & -0.10 & 0.255 & 0.0183 & 1.34              & 5.122 \\ \hline
\multirow{2}{*}{008366239} & 13.70\pm0.07 & $185.56\pm0.37$ & $88.2\pm1.3$ & $5239\pm170$ &       &       &        & $1.76\pm0.23$     &       \\
                           & 13.74          & 190.34          & 87.51        & 5142         & -0.22 & 0.255 & 0.0043 & 1.71              & 5.419 \\ \hline
\multirow{2}{*}{008475025} &  9.66\pm0.05 & $112.95\pm0.28$ & $74.8\pm1.4$ & $5056\pm154$ &       &       &        & $1.46\pm0.20$     &       \\
                           & 9.72           & 116.48          & 74.18        & 4751         & 0.08  & 0.240 & 0.0223 & 1.37              & 6.307 \\ \hline
\multirow{2}{*}{008718745} & 11.40\pm0.06 & $129.31\pm0.25$ & $79.4\pm1.3$ & $4825\pm167$ &       &       &        & $1.15\pm0.16$     &       \\
                           & 11.48          & 139.33          & 80.70        & 4681         & 0.14  & 0.230 & 0.0249 & 1.15              & 5.308 \\ \hline
\multirow{2}{*}{009267654} & 10.34\pm0.05 & $118.63\pm0.23$ & $78.4\pm1.4$ & $5029\pm156$ &       &       &        & $1.37\pm0.19$     &       \\
                           & 10.41          & 122.10          & 79.28        & 4794         & -0.04 & 0.230 & 0.0391 & 1.20              & 5.759 \\ \hline
\multirow{2}{*}{010257278} & 12.20\pm0.06 & $149.47\pm0.26$ & $79.8\pm1.4$ & $5055\pm152$ &       &       &        & $1.50\pm0.20$     &       \\
                           & 12.28          & 157.12          & 79.24        & 4981         & -0.16 & 0.235 & 0.0195 & 1.44              & 5.517 \\ \hline
\multirow{2}{*}{011353313} & 10.76\pm0.05 & $126.46\pm0.23$ & $76.0\pm1.4$ & $5198\pm168$ &       &       &        & $1.48\pm0.20$     &       \\
                           & 10.97          & 126.10          & 77.50        & 4941         & -0.16 & 0.275 & 0.0145 & 1.16              & 5.558 \\ \hline
\multirow{2}{*}{011913545} & 10.18\pm0.05 & $117.16\pm0.27$ & $77.8\pm1.3$ & $4845\pm145$ &       &       &        & $1.37\pm0.19$     &       \\
                           & 10.27          & 121.49          & 78.19        & 4637         & 0.20  & 0.230 & 0.0153 & 1.20              & 5.824 \\ \hline
\multirow{2}{*}{011968334} & 11.41\pm0.06 & $141.43\pm0.26$ & $78.1\pm1.4$ & $4914\pm144$ &       &       &        & $1.48\pm0.20$     &       \\
                           & 11.66          & 135.35          & 78.60        & 4831         & -0.10 & 0.235 & 0.0285 & 1.11              & 5.246 \\ \hline
\multirow{2}{*}{012008916} & 12.90\pm0.06 & $161.92\pm0.31$ & $80.5\pm1.3$ & $5002\pm170$ &       &       &        & $1.41\pm0.19$     &       \\
                           & 12.96          & 169.60          & 79.79        & 4739         & 0.20  & 0.240 & 0.0197 & 1.33              & 5.168 \\
\hline
\end{tabular}
\end{table*}

\section{Seismic modeling}
\label{seismic_mods}

From the original 19 young red giants studied by
\citet{corsaro2015}, we selected only those stars that
exhibit rotationally split dipole modes (triplets).
In some cases, depending on the stellar inclination angle,
some of the $l=1$ triplets were missing their central $m=0$ peaks 
although the split $m=\pm1$ components were clearly visible. 
In these cases we assumed a central $m=0$ component in the middle 
of the observed $m=\pm1$ frequencies, and we associated with it an 
error equal to three times the mean frequency error of the $l=0$ 
peaks (usually larger than the error on the $m=\pm1$ components). 
This choice is conservative given that the
asymmetry present in the full triplets (i.e., those that show central peaks) 
is usually smaller than this error.
Assuming the
presence of a central peak with this frequency uncertainty has virtually no effect
on the uncertainty on the inferred rotation rates, given that in these cases the splittings are computed simply as half
the distance in frequency of the $m=\pm1$ components, without involving the hypothetical
central component.

Equipped with these sets of pulsation frequencies, we set out to find
approximate seismic models for each target making use of the MESA 
stellar evolution suite \citep{paxton2011,paxton2013,paxton2015} together with
the GYRE pulsation code \citep{townsend2013}. The MESA suite includes the `astero' module,
which implements a downhill simplex search method \citep{nelder1965} to obtain the best stellar
parameters given a set of pulsation and spectroscopic data. To reduce computing
time during the search, we opted to include only the observed radial ($l=0$) modes and 
the dipole ($l=1$) modes. Including higher $l$ modes in the search,
 however desirable, would prohibitively increase
the time required to find suitable seismic models for all targets.

Our approach consisted of a combination of grid and downhill simplex
searches. We set up a grid 
of initial metallicities [Fe/H]$_\mathrm{ini}$, varying from -0.22 to 0.2 with 
0.06 steps (using a reference solar metallicity $Z_{\odot}/X_{\odot}$=0.02293 \citep{grevesse1998}) and
 initial helium content $Y_\mathrm{ini}$ varying from 0.23 to 0.3 
with 0.005 steps. For each pair ([Fe/H]$_\mathrm{ini}$ ,$Y_\mathrm{ini}$) 
we performed a downhill simplex search optimizing for 
initial mass and overshoot ($f_{\mathrm{ov}}$, expressed as a fraction
of the pressure scale height $H_P$), in addition 
to age. We have kept the mixing length parameter $\alpha_{MLT}$ fixed 
at its solar calibrated value of 1.9. We also used Eddington-gray atmospheres
and adopted the \citet{asplund2009} mixture together with OPAL opacity tables \citep{iglesias1996}. 
When computing mode frequencies, we peformed atmospheric corrections following
the method by \citet{kjeldsen2008} using a calibrated value of the exponent $b=4.81$ 
as reported by \citet{ball2014}. The exact value of $b$ is not critical as the observed modes
can still be identified one-to-one to model frequencies using slightly different values.
Results are summarized in Table~\ref{latabla}.

We performed a hare-and-hounds exercise to test our grid + downhill simplex approach.
For this we extracted mode frequencies from the best model of KIC007619745, added some noise
and used them as inputs to our search algorithm. The resulting model 
was satisfactorily close to
the original, especially regarding the rotational kernels derived from them. We
present more details in Appendix\,\ref{app}.

\FloatBarrier

\begin{figure*}[h!]
\begin{center}$
\begin{array}{cc}
\includegraphics[width=0.45\linewidth]{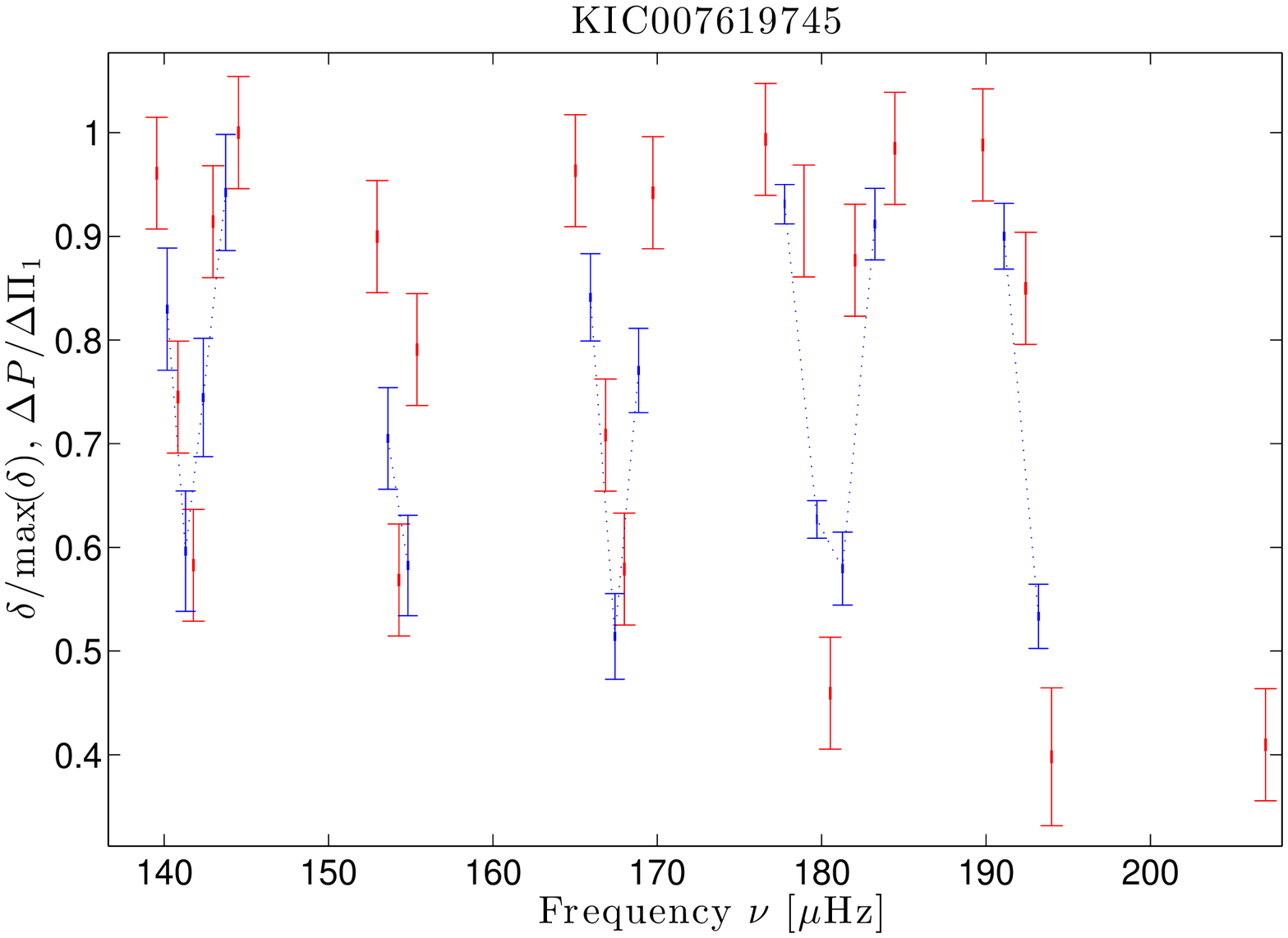}&
\includegraphics[width=0.45\linewidth]{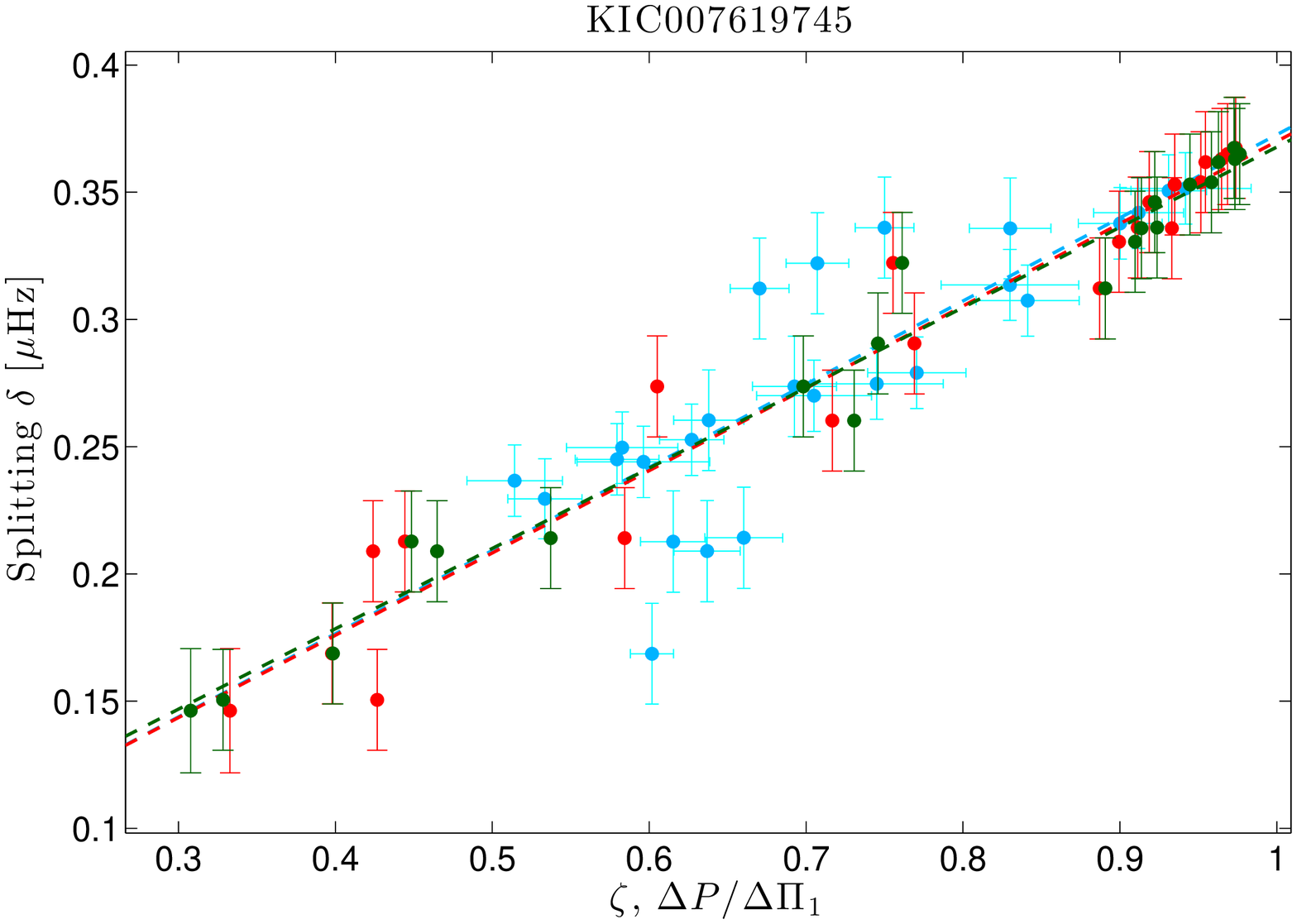}\\
\includegraphics[width=0.45\linewidth]{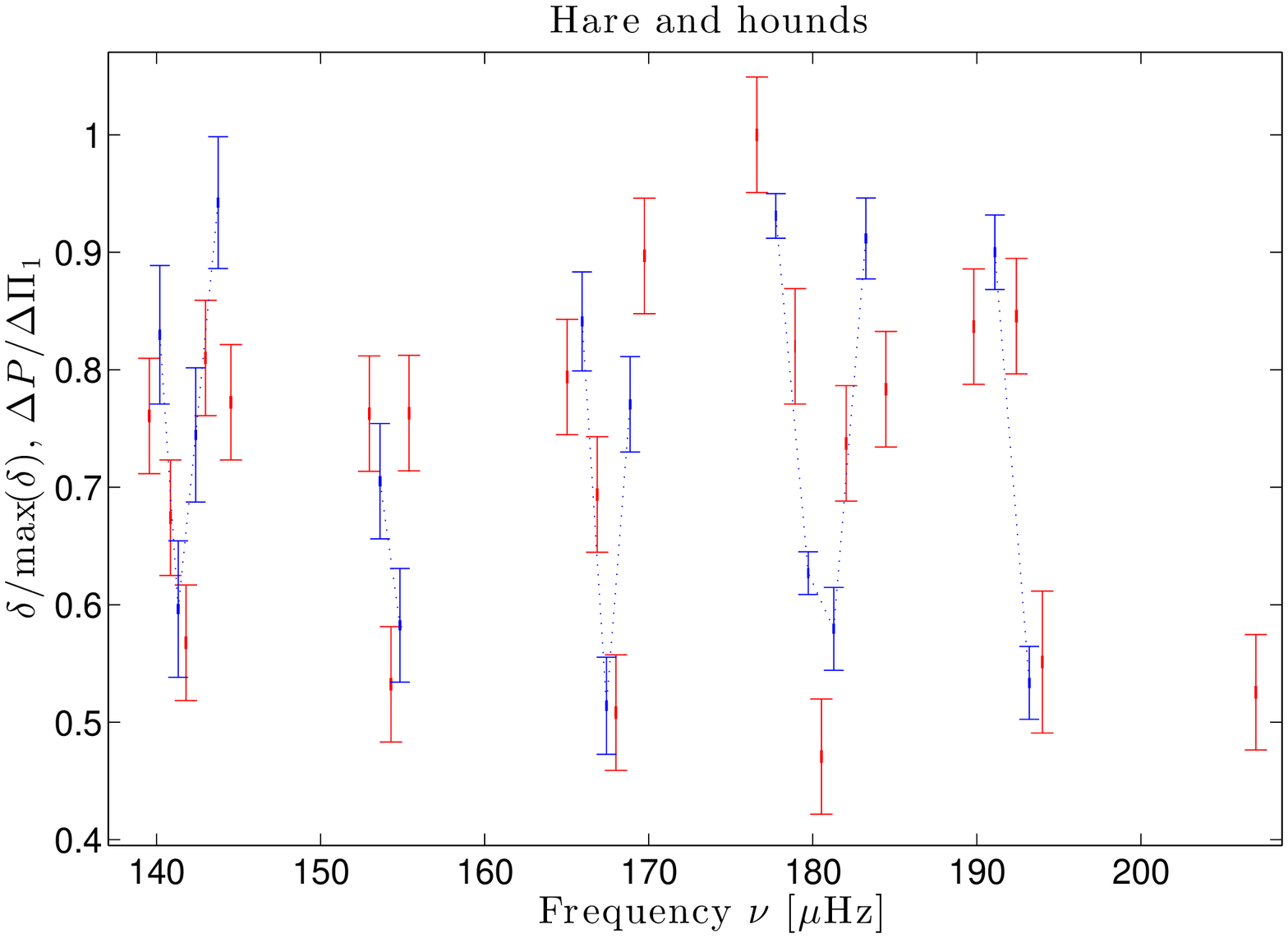}&
\includegraphics[width=0.45\linewidth]{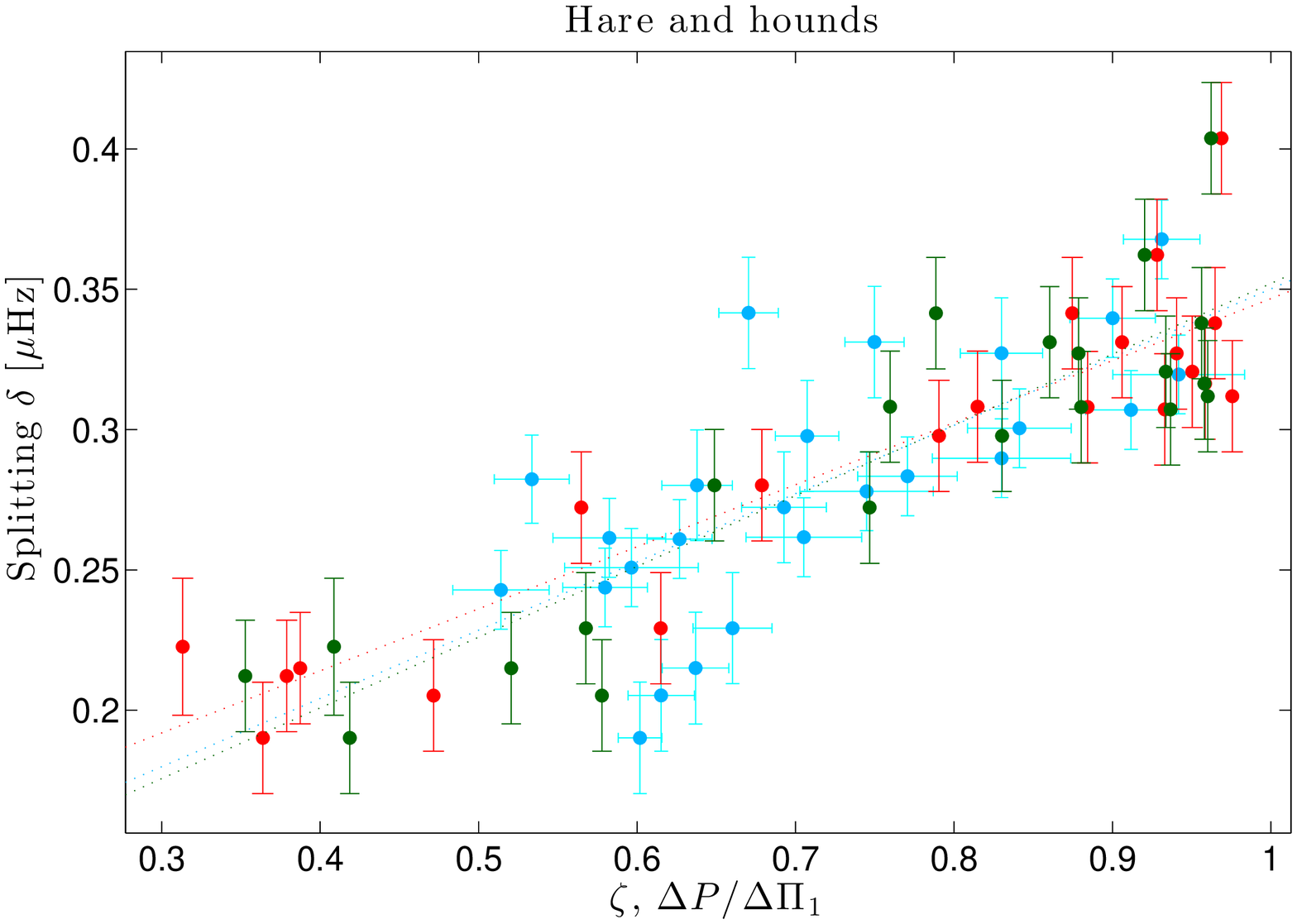}\\
\end{array}$
\end{center}
\caption{Left column plots show the observed rotational splittings, normalized
by their maximum value (red error bars)
and the observed period spacings, normalized by the observed period spacing 
$\Delta\Pi_1$ (blue error bars) as functions of mode frequency for the target KIC\,007619745 (top), and for the hare-and-hounds
exercise. Right column plots show the observed rotational splitting $\delta$ as a function of $\zeta_\mathrm{mod}$ (red
points and error bars), $\delta$ vs. $\zeta_\mathrm{as}$ (dark green points and error bars),
and both $\delta$ vs. $\Delta P_\nu/\Delta\Pi_1$ and $\delta_{\Delta P}$ vs. $\Delta P/\Delta\Pi_1$ 
(blue points and cyan error bars). See Section \ref{est_zeta}.}
\label{spp1}
\end{figure*}


\section{Internal rotation rates}
\label{inrot}

We estimated the internal rotation rates using two-zone models, Bayesian inference, and SOLA inversions
(all based on the kernels provided by the seismic models) in addition to the
model-independent method of \citet{goupil2013} as described in Section\,\ref{goupil}.
Figure\,\ref{spp1} illustrates the use of the trapping parameter and associated
linear fits as applied to the target KIC\,007619745 as an example, as well as
to the hare-and-hounds exercise explained in Appendix A.

We computed two-zone models with the inter-zone 
boundary in the middle of the evanescent zone for each target. For the SOLA inversions we set the
trade-off parameter $\mu$ to zero since it made negligible difference on the results.
In addition, for SOLA, we computed rotation rates at two radial locations, one at $r/R_*=10^{-3}$, deep in the
radiative cores, and the other at the surface, $r/R_*=1$. The estimates of the 
rotation rates through SOLA do depend slightly on the width $s$ of the target functions
used. To select the optimal value of $s,$ we took the two-zone rotation rates determined
earlier and computed their associated splittings via Eq.\,\ref{forward}. Then, using these
synthetic splittings, we iteratively determined the optimal $s$ values required to
exactly reproduce the two-zone rotation rates. With the widths $s$ thus determined,
we then computed the rotation rates at $r/R_*=10^{-3}$ and at $r/R_*=1$ using the observed
rotational splittings. The rotation rates estimated in this way are presented in 
Table\,\ref{omtabla} and in Figs.\,\ref{omg1} and \ref{omg2}.

\begin{table*}
\caption{Rotation rates in nHz estimated from two-zone models, SOLA inversions, Bayesian inference,
 linear fits of $\delta$ vs $\zeta_\mathrm{mod}$, linear fits of $\delta$ vs $\zeta_\mathrm{as}$, and linear 
 fits of $\delta$ vs $\Delta P/\Delta \Pi_1$. For each target,
 the top row is the rotation rate of the core, and the bottom row is the rotation rate of the envelope. The
 four leftmost columns correspond to rotation rates based on seismic models, the next two columns present model-independent
 rotation averages. The last column
 presents an overall rotation rate computed as the mean from the six methods used; the corresponding 
 overall error is computed by adding the mean variance to the \emph{inter-}variance across the methods. See also Figs\,\ref{omg1} and \ref{omg2}}
\label{omtabla}
\scriptsize     
\centering                          
\begin{tabular}{c D{.}{.}{5.7}  D{.}{.}{5.7}  D{.}{.}{5.7}  D{.}{.}{5.7}  D{.}{.}{5.7}  D{.}{.}{5.7}  D{.}{.}{5.7} }        
\hline\hline                  
\multicolumn{1}{c}{KIC id}      & \multicolumn{1}{c}{Two-zone} 
                                                        & \multicolumn{1}{c}{SOLA} 
                                                        & \multicolumn{1}{c}{Bayes} 
                                                        & \multicolumn{1}{c}{$\delta(\zeta_\mathrm{mod})$}
                                                        & \multicolumn{1}{c}{$\delta(\zeta_\mathrm{as})$}
                                                        & \multicolumn{1}{c}{$\delta(\Delta P/\Delta\Pi_1)$}
                                                        & \multicolumn{1}{c}{Overall}\\
\hline
\multirow{2}{*}{003744043}      & 541.8 \pm 9.2 & 525.9 \pm 15.6 & 544.8 \pm 9.3 & 537.1 \pm 16.0 & 536.9 \pm 16.2 & 529.6 \pm 17.9 & 536.0 \pm 16.1 \\
                                                        & 75.8 \pm 25.9 & 110.9 \pm 29.8 & 72.6 \pm 25.2 & 88.0 \pm 28.2 & 78.9 \pm 30.7 & 92.0 \pm 32.2 & 86.4 \pm 32.1 \\ \hline
                                                        
\multirow{2}{*}{006144777}      & 476.9 \pm 2.9 & 458.6 \pm 8.9 & 479.0 \pm 2.7 & 470.5 \pm 7.9 & 485.9 \pm 5.8 & 485.3 \pm 12.3 & 476.0 \pm 12.7 \\ 
                                                        & 3.0 \pm 10.4 & 73.2 \pm 19.1 & 8.0 \pm 8.0 & 48.3 \pm 14.8 & 6.8 \pm 16.2 & 26.5 \pm 25.4 & 27.6 \pm 32.6 \\ \hline
                                                        
\multirow{2}{*}{007060732}  & 629.0 \pm 2.1 & 620.7 \pm 15.1 & 633.0 \pm 2.2 & 630.9 \pm 12.6 & 649.6 \pm 12.7 & 646.7 \pm 20.1 & 635.0 \pm 16.8 \\
                                                        & 66.1 \pm 20.4 & 74.9 \pm 39.4 & 64.1 \pm 19.9 & 62.1 \pm 35.2 & 4.6 \pm 39.5 & 29.3 \pm 39.9 & 50.2 \pm 43.2 \\

\multirow{2}{*}{007619745}      & 731.8 \pm 12.8 & 733.3 \pm 14.0 & 738.1 \pm 13.0 & 740.0 \pm 13.1 & 735.6 \pm 12.8 & 745.2 \pm 22.8 & 737.3 \pm 16.0 \\ 
                                                        & 50.9 \pm 16.1 & 48.1 \pm 18.0 & 48.2 \pm 15.8 & 46.5 \pm 16.0 & 52.2 \pm 15.5 & 45.8 \pm 24.5 & 48.6 \pm 18.1 \\ \hline

\multirow{2}{*}{008366239}      & 434.3 \pm 17.9 & 436.5 \pm 19.3 & 439.1 \pm 18.2 & 440.0 \pm 18.3 & 461.5 \pm 20.2 & 432.1 \pm 28.4 & 440.6 \pm 23.3 \\ 
                                                        & 80.1 \pm 20.4 & 80.2 \pm 22.6 & 79.6 \pm 19.9 & 79.2 \pm 20.0 & 36.7 \pm 25.3 & 90.2 \pm 23.5 & 74.3 \pm 29.0 \\ \hline

\multirow{2}{*}{008475025}      & 622.6 \pm 0.5 & 607.4 \pm 3.3 & 625.2 \pm 0.5 & 617.9 \pm 2.5 & 660.9 \pm 2.4 & 622.7 \pm 19.3 & 626.1 \pm 19.9 \\ 
                                                        & 102.1 \pm 1.2 & 130.3 \pm 6.7 & 98.7 \pm 1.1 & 115.5 \pm 6.0 & 15.8 \pm 6.8 & 76.4 \pm 27.7 & 89.8 \pm 42.3 \\ \hline

\multirow{2}{*}{008718745}      & 813.7 \pm 40.1 & 749.0 \pm 37.7 & 816.7 \pm 40.4 & 812.6 \pm 40.7 & 853.7 \pm 43.8 & 803.5 \pm 41.7 & 808.2 \pm 53.0 \\ 
                                                        & 131.9 \pm 37.8 & 177.9 \pm 44.9 & 128.3 \pm 36.8 & 151.3 \pm 39.2 & 36.4 \pm 52.2 & 80.9 \pm 52.5 & 117.8 \pm 67.7 \\ \hline

\multirow{2}{*}{009267654}      & 911.4 \pm 0.8 & 908.2 \pm 3.1 & 916.2 \pm 0.8 & 915.3 \pm 2.1 & 937.3 \pm 2.4 & 939.4 \pm 19.3 & 921.3 \pm 15.8 \\ 
                                                        & 54.6 \pm 5.4 & 65.0 \pm 9.8 & 52.9 \pm 5.3 & 57.9 \pm 8.4 & -23.1 \pm 10.1 & 20.8 \pm 21.8 & 38.0 \pm 35.6 \\ \hline

\multirow{2}{*}{010257278}      & 959.6 \pm 33.6 & 930.6 \pm 36.4 & 966.9 \pm 34.0 & 969.0 \pm 34.1 & 1024.2 \pm 38.7 & 1002.3 \pm 49.3 & 975.4 \pm 50.4 \\ 
                                                        & 137.7 \pm 49.0 & 188.3 \pm 55.9 & 132.9 \pm 48.0 & 131.8 \pm 48.0 & -1.5 \pm 63.4 & 33.6 \pm 64.9 & 103.8 \pm 90.8 \\ \hline

\multirow{2}{*}{011353313}      & 891.3 \pm 23.0 & 901.2 \pm 25.4 & 888.3 \pm 21.5 & 901.1 \pm 25.6 & 914.4 \pm 26.9 & 908.2 \pm 37.5 & 900.7 \pm 28.9 \\ 
                                                        & 78.7 \pm 27.5 & 88.9 \pm 32.9 & 77.0 \pm 26.2 & 86.5 \pm 35.5 & 50.7 \pm 40.4 & 77.6 \pm 51.5 & 76.6 \pm 39.1 \\ \hline

\multirow{2}{*}{011913545}      & 559.1 \pm 0.4 & 544.7 \pm 2.6 & 559.2 \pm 0.3 & 550.7 \pm 2.2 & 566.1 \pm 2.6 & 559.5 \pm 11.6 & 556.5 \pm 9.1 \\ 
                                                        & -13.7 \pm 2.5 & 53.1 \pm 7.9 & 0.4 \pm 0.4 & 45.3 \pm 7.1 & -3.9 \pm 9.1 & 22.4 \pm 14.0 & 17.3 \pm 28.7 \\ \hline

\multirow{2}{*}{011968334}      & 767.8 \pm 9.9 & 727.8 \pm 23.8 & 767.6 \pm 9.9 & 729.5 \pm 22.2 & 746.6 \pm 21.6 & 743.0 \pm 25.6 & 747.0 \pm 26.5 \\ 
                                                        & 119.4 \pm 15.2 & 73.6 \pm 27.1 & 127.0 \pm 14.2 & 74.2 \pm 28.1 & 22.7 \pm 36.1 & 49.0 \pm 34.1 & 77.6 \pm 48.4 \\ \hline

\multirow{2}{*}{012008916}      & 815.5 \pm 0.9 & 735.5 \pm 4.2 & 822.0 \pm 0.9 & 745.9 \pm 3.7 & 776.7 \pm 4.7 & 712.2 \pm 16.1 & 767.9 \pm 45.1 \\ 
                                                        & 10.4 \pm 3.2 & 113.6 \pm 7.2 & 10.1 \pm 3.1 & 108.8 \pm 6.2 & 45.4 \pm 7.7 & 80.3 \pm 15.1 & 61.5 \pm 47.2 \\ \hline
\hline  
\end{tabular}
\end{table*}

We assume that the observational errors on the splittings
and on the mode frequencies are normally distributed. Since in general, a given mode
frequency has two period spacings associated with it and each period spacing is associated with
two mode frequencies (and therefore two rotational splittings), we opted to take interpolated
values as explained in Section~\ref{est_zeta}. We performed
linear fits to the resulting set to estimate the average rotation rates
of the core $\Omega_g$ and the envelope $\Omega_p$ according to Eq.\,\ref{goupil_eq}.
A straightforward Monte Carlo approach using the observed rotational splittings and frequencies,
together with their normally distributed errors, reveals a correlation between $\Omega_g$ 
and $\Omega_p$ that is a reflection of the fact that in Eq.\,\ref{goupil_eq}
the slope and the intercept are not independent of each other.
The rotation rates using this technique are shown in Figs.\,\ref{omg1} and \ref{omg2} in the Appendix B as a cloud of small
violet circles, the blue box is centered on the mean, and its size corresponds to the 1$\sigma$
 standard deviation of the points in the cloud.

Table \ref{omtabla} summarizes the rotation rates for all 13 targets obtained by all the methods described in this work.


\section{Discussion}
\label{disc}

\begin{figure*}[h!]
\begin{center}$
\begin{array}{cc}
\includegraphics[width=0.45\linewidth]{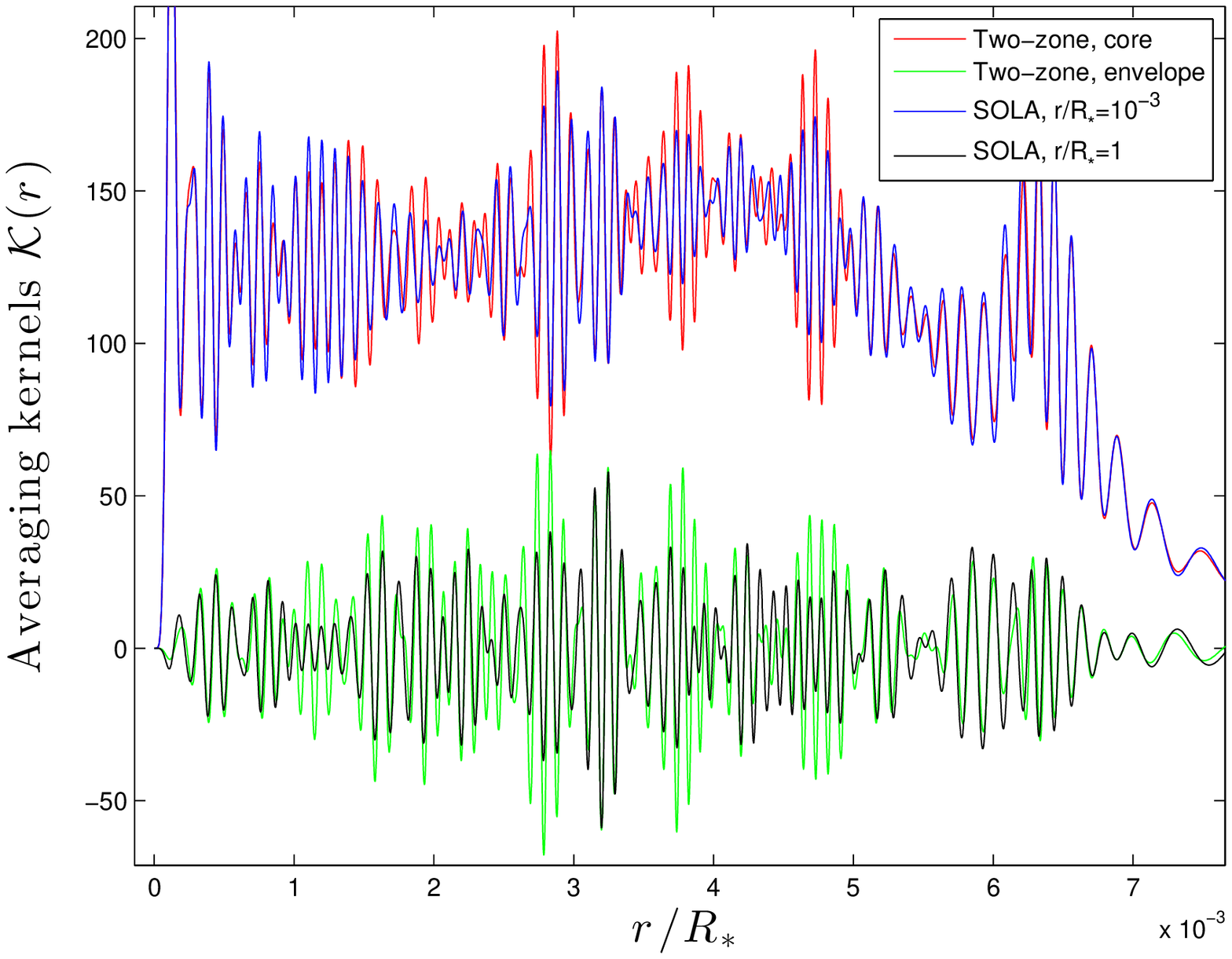}&
\includegraphics[width=0.45\linewidth]{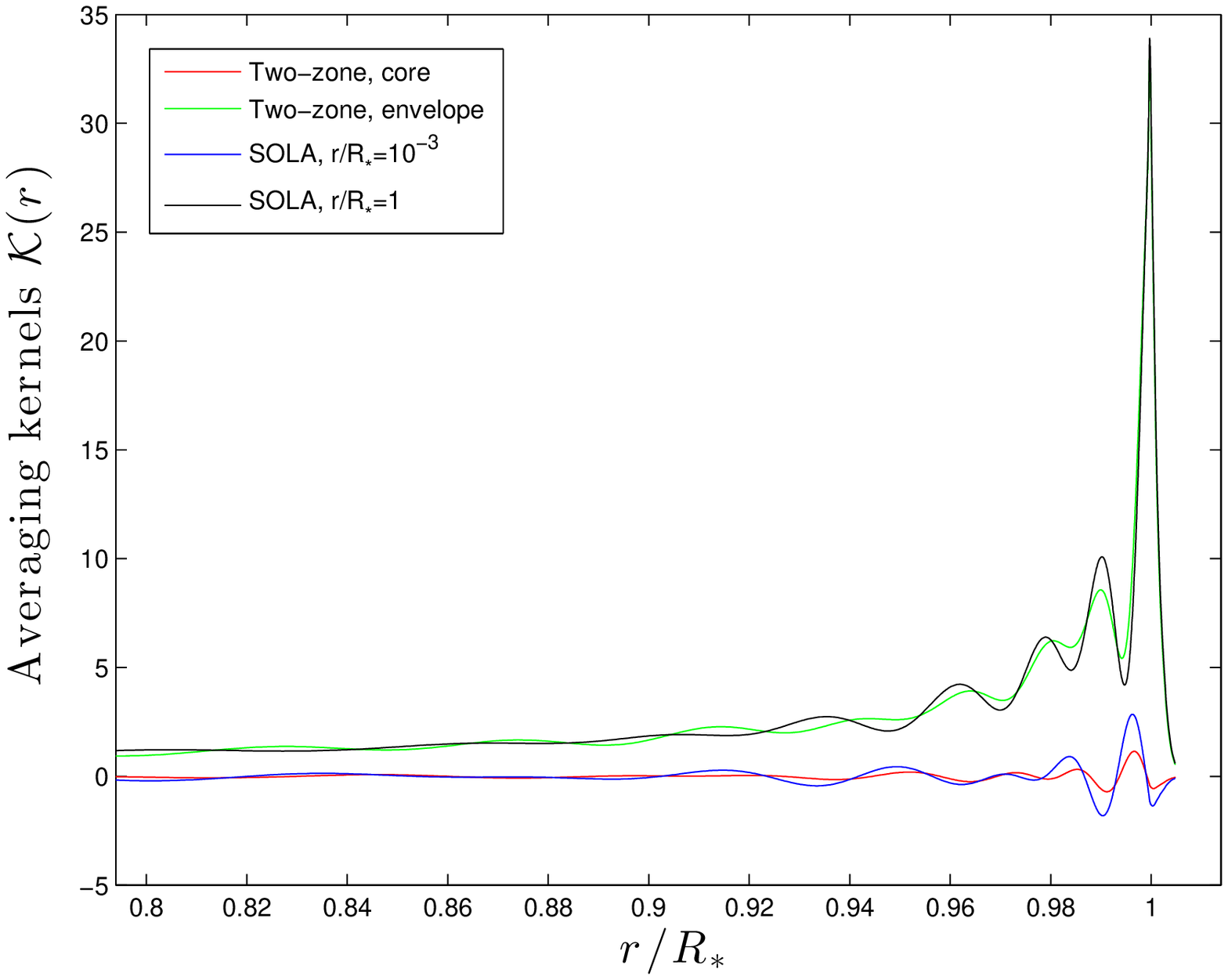}\\
\end{array}$
\end{center}
\caption{Averaging kernels $\mathcal{K}(r)$ for both SOLA
inversions at $r/R_*=10^{-3}$, $r/R_*=1$ and a two-zone model
with $x_c=0.1$ (from the best model of KIC007619745). The left panel
shows the central region, and the right panel shows the surface region.}
\label{avker}
\end{figure*}

The rotation rates for most of the targets show good agreement, while a few present some scatter
in the envelope averages. The rotation rates agree with each other within 2$\sigma$ except for one case (see further below).

There are minor differences in the ideal case of many exactly measured splittings and exact
seismic models, as described in Section~\ref{tests}. The differences in this case are attributed to the different
nature of the averages as computed from each method, for example, the averages as defined by Eqs.~\ref{avtur1} and \ref{avtur2}
are not exactly the same as the averages from Eq.~\ref{eq:avker}, even if we had many kernels $K(r)$ at our disposal to have
very well localized averaging kernels $\mathcal{K}(r',r)$.

However, considering the scatter of the predicted averages, we can still constrain the rotation rates in our targets with
an error of about $0.05~\mu$Hz except in a few cases. No more than two averages that are well resolved spatially could be obtained
using inversions and the seismic models, which was our hope given the previous success reported by \citet{dimauro2016}.

\begin{figure}
\begin{center}
\includegraphics[width=1.0\linewidth]{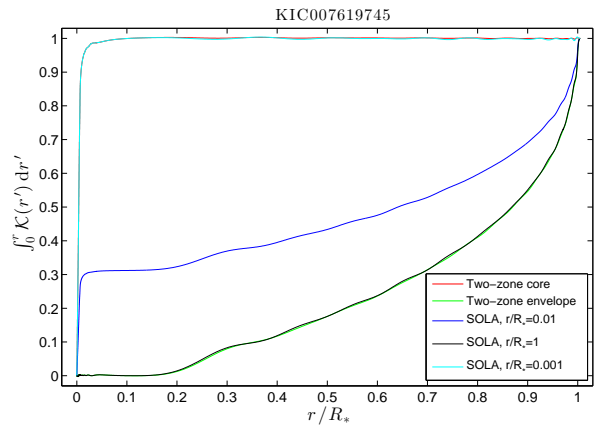}
\caption{ Cumulative integrals of the averaging kernels $\mathcal{K}(r)$ 
(from the best model of KIC007619745). The integrals corresponding to the inner two-zone
model and SOLA at $r/R_*=10^{-3}$ are almost identical, the same
holds for the outer two-zone
and SOLA at $r/R_*=1$. A SOLA kernel at $r/R_*=0.01$ shows already significant leakage from other 
regions of the star, as shown by its cummulative integral not reaching unity anywhere close to that location.}
\label{cumm}
\end{center}
\end{figure}

We have chosen one of the stars in the sample (KIC\,007619745) as an illustrative case to explain
why only two values could be obtained. In Fig.\,\ref{avker} we show the averaging kernels $\mathcal{K}(r)$ for the two-zone
model (inner and outer zones) and SOLA (at $r/R_*=10^{-3}$ and $r/R_*=1$). According
to Eq.\,\ref{eq:avker}, these are simply the weight functions involved in the average.
They are localized (at least enough to resolve the core and the envelope), 
with the inner-zone kernel and the $r/R_*=10^{-3}$ SOLA kernel
concentrating in approximately the same region in the stellar core. On the envelope,
the outer-zone kernel and the $r/R_*=1$ SOLA kernel coincide roughly as well, while they
oscillate rapidly around zero as the radius is varied in the core regions. These
rapid oscillations around zero do not present a problem since we are only interested
in the integrated averages. Indeed, as Fig.\,\ref{cumm} shows, the cumulative integrals
of the averaging kernels are fortunately insensitive to the rapid oscillations. The
outer-zone and the $r/R_*=1$ SOLA cumulative kernels start
growing only at around $r/R_*=0.2$, where the inner-zone and the $r/R_*=10^{-3}$
 cumulative kernels have already reached unity. These properties allow us to estimate
rotation rate averages of the core separately from the envelope. The inner zone averaging kernel in the SOLA inversions is basically the same when we use other radial
locations well inside the radiative core, that is to say, we obtain the same average. As we approach the hydrogen burning region
near $r/R_*=0.01,$ however, there is significant contamination from the outer regions of the star, as is shown
in Fig.~\ref{cumm}. All of our targets exhibit essentially identical behavior. 

\begin{figure*}
\begin{center}$
\begin{array}{cc}
\includegraphics[width=0.45\linewidth]{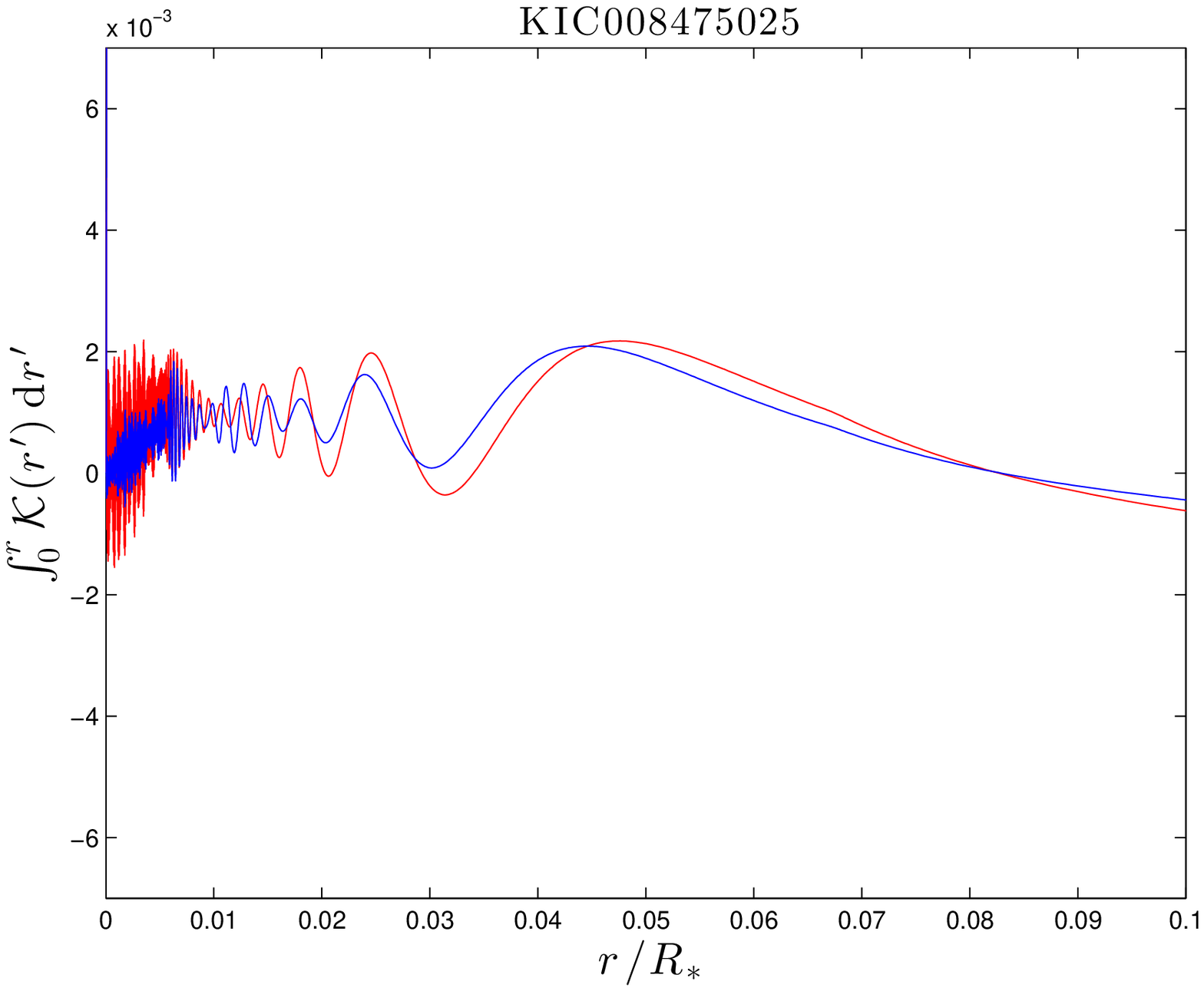}&
\includegraphics[width=0.45\linewidth]{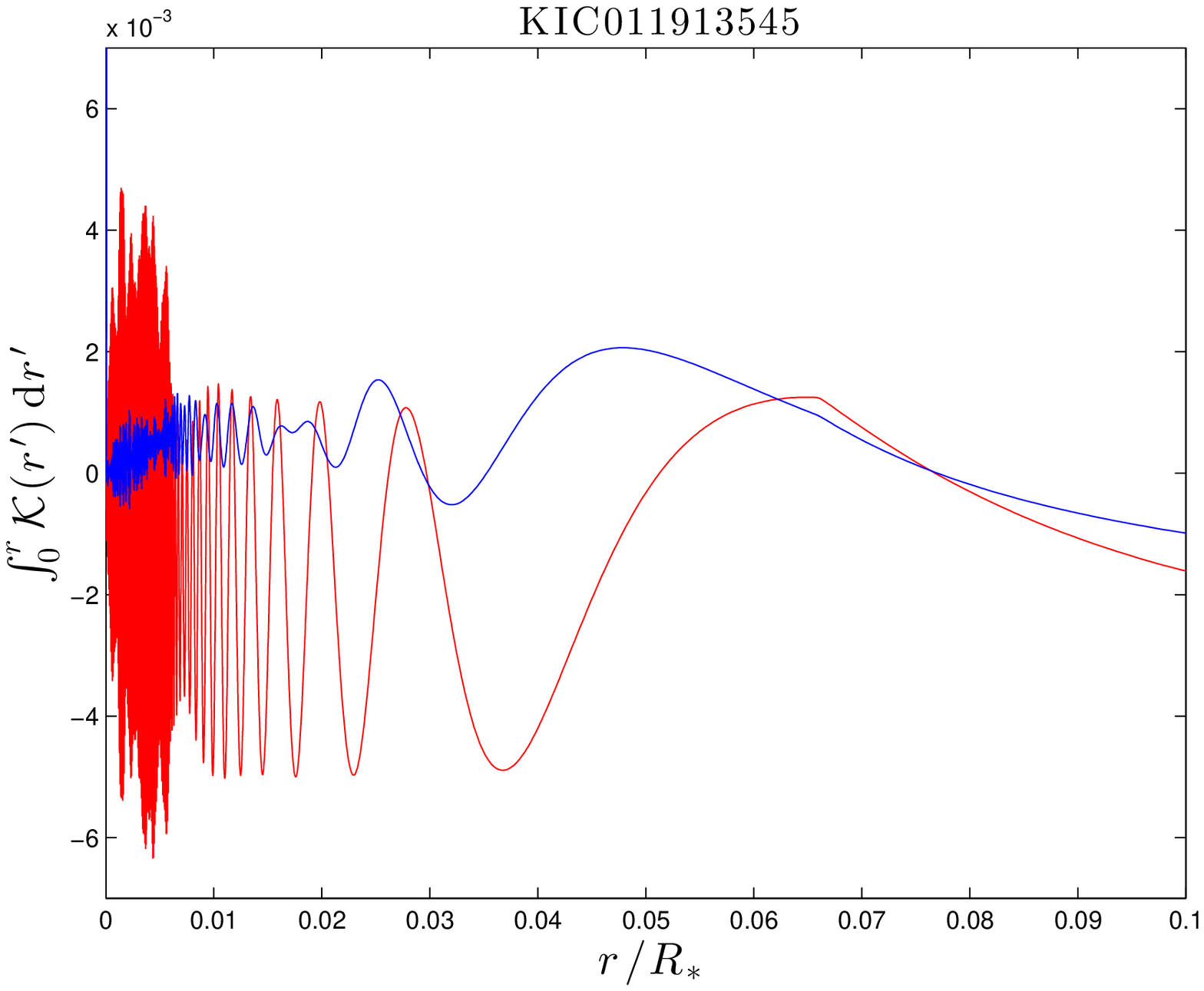}\\
\end{array}$
\end{center}
\caption{ Cumulative integral of averaging kernels $\mathcal{K}(r)$ from SOLA
inversions at $r/R_*=1$ (blue curves) and the cumulative integral
from the outer two-zone model kernels (red curves). The left panel shows
the case of KIC\,008475025, where both cumulative integrals mostly coincide, are close to zero, and lead to similar
average rotation rates of the envelope without appreciable leakage from central regions. In the case of KIC\,011913545 (right panel) we see a
difference between the cumulative integrals, particularly, the red curve is consistently below zero in the core region, 
which leads to leakage from these regions affecting the envelope averages.  
}
\label{goodbad}
\end{figure*}

 We note, however, that
although the two-zone and SOLA kernels coincide for most targets, there were some instances where they differ
slightly. In Fig.\,\ref{goodbad} we show an example illustrating this point. The two-zone kernels exhibit some
leakage that leads to appreciable differences in the rotation averages when compared with the SOLA kernels, for instance.
This is the case for targets KIC\,011913545 and KIC\,012008916.

\begin{figure}
\begin{center}$
\begin{array}{cc}
\includegraphics[width=0.9\linewidth]{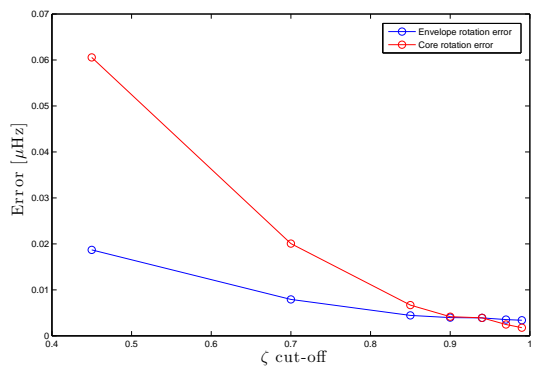}&
\end{array}$
\end{center}
\caption{ Error on the rotation averages using a synthetic rotation profile. Starting with the seismic model of
KIC\,012008916, we selected all the modes that had a trapping parameter $\zeta$ smaller than a particular cut-off.
The total number of modes in the model is 45 (cutoff $\zeta=1$). We computed two-zone inversions using synthetic splittings from modes with
trapping parameters smaller than a given $\zeta$ cutoff.  More splittings used certainly improve the error
on the predicted rotation averages, but this does not take into account any systematic errors that may be present.  }
\label{final_test}
\end{figure}

The larger the number of splittings observed, the better the localization properties of the kernels will be
and the error on the predicted rotation rates will be lower, see Fig.\ref{final_test}. We note that the predicted errors as computed from
Eq.\,\ref{eq:variance} \emph{\textup{do not}} include any systematic errors arising from unmodeled physics, which means
that the predicted rotation rates may be precise, but not accurate. This certainly constitutes a source for discrepancies 
in addition to the variability induced by poorly localized averaging kernels, as discussed above.

Accidental mode misidentification can potentially lead to an inadequate seismic model. As noted earlier,
we are considering only modes with a detection probability higher than 0.99 \citep{corsaro2015}. In principle, we therefore do not expect complications arising from this issue. Since a misidentified
dipole mode might reveal itself as an outlier if its rotational splitting differs considerably from its prediction, we closely
examined the
rotational splittings of the star KIC\,012008916 and compared them with the predicted two-zone model splittings.
However, no particular mode stands out in a clear way. Still, we could prune out the modes whose splittings differ the most
from the two-zone prediction. This exercise modifies the predicted rotation averages and removes
the scatter bringing the predictions across the six methods in good agreement. This is still not satisfactory, however,
since we have no means to a priori justify \textup{}the adequacy of a two-zone rotation profile, or of any other particular profile, for that matter.

To conclude our discussion, we would like to point out that the target KIC\,12008916 is coincidentally one of the few in the sample that has a very high inclination angle, that is, no
central peaks were detected for it. This could be connected to the scatter in the rotation rates in some way.

\section{Summary and conclusion}
\label{conc}

Building upon the mode-fitting and identification work of \citet{corsaro2015}, who analyzed the power spectra of 19 red giant targets 
recorded by the \emph{Kepler} space telescope, we have estimated
the average core and envelope rotation rate of 13 of these targets that showed
clear rotationally split mixed modes. We employed the
model-independent method developed originally by \citet{goupil2013} and improved later by 
\citet{deheuvels2015} and \citet{mosser2015}. This model-independent method aims 
to provide two rotational averages, one for the \textup{\textit{g}}-mode cavity and another for the \textit{p}-mode cavity.
To investigate the possibility of obtaining more detailed information about the rotation profile,
we have used the traditional approach
of searching for optimal seismic models with the aid of the MESA stellar evolution
suite and the GYRE oscillation code. 

We used a total of six different methods to compute the averages, four of them based on the optimal seismic models:
SOLA inversions, two-zone inversions, Bayesian inference on two-zone models,
 and linear fits of the rotational splittings $\delta$ as functions of the 
trapping parameter $\zeta_\mathrm{mod}$. We also used the model-independent method as implemented
by \citet{deheuvels2015} and a variation of it based on the result by \citet{mosser2015}.

Before we applied these methods to our sample of red giants, we took a particular seismic model (that of KIC007619745) as a
`true' reference model, and in conjunction with six synthetic rotation profiles, we proceeded to compute
the associated rotational splittings. Using these as inputs, we compared the predictions from all the methods
against the 'true' synthetic rotation profiles. We found good agreement in general, with some small
differences that can be attributed to the slightly different nature of the averages produced by each method. 

All the rotational kernels of our 13 targets allow the computation of two
distinct rotation rates. We used the functional form of the rotation profiles used in the study
by \citet{klion2016} to test whether we could
determine the approximate location in the star where the differential rotation takes place. The results were
negative given the magnitude of the observational errors. The information contained in the splittings
is insufficient to localize the region with differential rotation. This is perhaps closely
related to the fact that no good averaging kernels could be found for intermediate regions of any of the targets.

The averages that we could obtain, however, are in good agreement with the average rotation rates of the
\textit{g}- and \textit{p}-mode cavities as computed from the model-independent method for most of the targets, while a few targets
present more scatter in the predicted averages across the different methods.  The rotation rates agree
with each other within 2$\sigma$ with only one exception, that of the star KIC\,012008916. We identify
the poorly localized two-zone averaging kernels as a contributor to the discrepancy. However,
many sources of systematic errors exist that cannot be ruled out. There are indeed many
simplifying assumptions involved in the stellar evolution code that directly affect the mode kernels, which means
that some seismic models may not be adequate. In this respect, the model-independent methods are
more reliable since they are essentially free of such systematic errors.

At any rate, the results for the rotation of the radiative cores show better agreement
across all methods than the rotation of the envelopes, which is related to the fact that
for \textit{p}-dominated mixed modes the trapping parameter is $\zeta\approx0.5$, far from $0$. The cores in this
target selection are spinning about 5 to 10 times faster than their envelopes,
which is consistent with previous studies and still calls for a better
understanding of the angular momentum redistribution in stars in the red
giant branch.

\begin{acknowledgements}
The authors express their gratitude to the referee, whose comments greatly improved
the manuscript. S.~T. would like to thank Ehsan Moravveji for enlightening discussions and help with
the MESA code. S.~T. received partial funding from the ERC Advanced Grant ROTANUT (No. 670874).
 E.~C. acknowledges funding by the European Community’s Seventh
Framework Programme (FP7/2007-2013) under grant agreement No. 312844 (SPACEINN).
R.~A.~G acknowledges the support from the CNES GOLF and PLATO grants at CEA 
as well as the ANR (Agence Nationale de la Recherche, France) program 
IDEE (n ANR-12-BS05-0008) “Interaction Des \'Etoiles et des Exoplan\`etes”.
We are grateful to the {\it Kepler\/} team and everybody who has
contributed to making this mission possible. Funding for the {\it Kepler\/}
Mission was provided by NASA's Science Mission Directorate.  The research
leading to these results has received funding from the Fund for Scientific
Research of Flanders (FWO) under project O6260.
\end{acknowledgements}

\bibliographystyle{aa} 
\bibliography{thirteen} 

\begin{thebibliography}{44}
\expandafter\ifx\csname natexlab\endcsname\relax\def\natexlab#1{#1}\fi

\bibitem[{{Aerts} {et~al.}(2010){Aerts}, {Christensen-Dalsgaard}, \&
  {Kurtz}}]{Aerts2010asteroseismology}
{Aerts}, C., {Christensen-Dalsgaard}, J., \& {Kurtz}, D.~W. 2010,
  Asteroseismology, Astronomy and Astrophysics Library (Springer Heidelberg)

\bibitem[{{Alvan} {et~al.}(2013){Alvan}, {Mathis}, \& {Decressin}}]{alvan2013}
{Alvan}, L., {Mathis}, S., \& {Decressin}, T. 2013, \aap, 553, A86

\bibitem[{Asplund {et~al.}(2009)Asplund, Grevesse, Sauval, \&
  Scott}]{asplund2009}
Asplund, M., Grevesse, N., Sauval, A.~J., \& Scott, P. 2009, arXiv preprint
  arXiv:0909.0948

\bibitem[{Ball \& Gizon(2014)}]{ball2014}
Ball, W.~H. \& Gizon, L. 2014, Astronomy \& Astrophysics, 568, A123

\bibitem[{{Beck} {et~al.}(2011){Beck}, {Bedding}, {Mosser}, {Stello}, {Garcia},
  {Kallinger}, {Hekker}, {Elsworth}, {Frandsen}, {Carrier}, {De Ridder},
  {Aerts}, {White}, {Huber}, {Dupret}, {Montalb{\'a}n}, {Miglio}, {Noels},
  {Chaplin}, {Kjeldsen}, {Christensen-Dalsgaard}, {Gilliland}, {Brown},
  {Kawaler}, {Mathur}, \& {Jenkins}}]{beck2011}
{Beck}, P.~G., {Bedding}, T.~R., {Mosser}, B., {et~al.} 2011, Science, 332, 205

\bibitem[{{Beck} {et~al.}(2012){Beck}, {Montalban}, {Kallinger}, {De Ridder},
  {Aerts}, {Garc{\'{\i}}a}, {Hekker}, {Dupret}, {Mosser}, {Eggenberger},
  {Stello}, {Elsworth}, {Frandsen}, {Carrier}, {Hillen}, {Gruberbauer},
  {Christensen-Dalsgaard}, {Miglio}, {Valentini}, {Bedding}, {Kjeldsen},
  {Girouard}, {Hall}, \& {Ibrahim}}]{beck2012}
{Beck}, P.~G., {Montalban}, J., {Kallinger}, T., {et~al.} 2012, \nat, 481, 55

\bibitem[{{Belkacem, K.} {et~al.}(2015){Belkacem, K.}, {Marques, J. P.},
  {Goupil, M. J.}, {Sonoi, T.}, {Ouazzani, R. M.}, {Dupret, M. A.}, {Mathis,
  S.}, {Mosser, B.}, \& {Grosjean, M.}}]{belkacem2015}
{Belkacem, K.}, {Marques, J. P.}, {Goupil, M. J.}, {et~al.} 2015, \aap, 579,
  A30

\bibitem[{Borucki {et~al.}(2010)Borucki, Koch, Basri, Batalha, Brown, Caldwell,
  Caldwell, Christensen-Dalsgaard, Cochran, DeVore, Dunham, Dupree, Gautier,
  Geary, Gilliland, Gould, Howell, Jenkins, Kondo, Latham, Marcy, Meibom,
  Kjeldsen, Lissauer, Monet, Morrison, Sasselov, Tarter, Boss, Brownlee, Owen,
  Buzasi, Charbonneau, Doyle, Fortney, Ford, Holman, Seager, Steffen, Welsh,
  Rowe, Anderson, Buchhave, Ciardi, Walkowicz, Sherry, Horch, Isaacson,
  Everett, Fischer, Torres, Johnson, Endl, MacQueen, Bryson, Dotson, Haas,
  Kolodziejczak, Van~Cleve, Chandrasekaran, Twicken, Quintana, Clarke, Allen,
  Li, Wu, Tenenbaum, Verner, Bruhweiler, Barnes, \& Prsa}]{borucki2010}
Borucki, W.~J., Koch, D., Basri, G., {et~al.} 2010, Science, 327, 977

\bibitem[{{Cantiello} {et~al.}(2014){Cantiello}, {Mankovich}, {Bildsten},
  {Christensen-Dalsgaard}, \& {Paxton}}]{cantiello2014}
{Cantiello}, M., {Mankovich}, C., {Bildsten}, L., {Christensen-Dalsgaard}, J.,
  \& {Paxton}, B. 2014, \apj, 788, 93

\bibitem[{Corsaro {et~al.}(2015)Corsaro, De~Ridder, \&
  Garc{\'\i}a}]{corsaro2015}
Corsaro, E., De~Ridder, J., \& Garc{\'\i}a, R. 2015, Astronomy \& Astrophysics,
  579, A83

\bibitem[{{Corsaro} {et~al.}(2013){Corsaro}, {Fr{\"o}hlich}, {Bonanno},
  {Huber}, {Bedding}, {Benomar}, {De Ridder}, \& {Stello}}]{corsaro2013}
{Corsaro}, E., {Fr{\"o}hlich}, H.-E., {Bonanno}, A., {et~al.} 2013, \mnras,
  430, 2313

\bibitem[{{Corsaro, E.} \& {De Ridder, J.}(2014)}]{corsaro2014}
{Corsaro, E.} \& {De Ridder, J.} 2014, \aap, 571, A71

\bibitem[{{De Ridder} {et~al.}(2009){De Ridder}, {Barban}, {Baudin}, {Carrier},
  {Hatzes}, {Hekker}, {Kallinger}, {Weiss}, {Baglin}, {Auvergne}, {Samadi},
  {Barge}, \& {Deleuil}}]{deridder2009}
{De Ridder}, J., {Barban}, C., {Baudin}, F., {et~al.} 2009, \nat, 459, 398

\bibitem[{{Deheuvels} {et~al.}(2015){Deheuvels}, {Ballot}, {Beck}, {Mosser},
  {{\O}stensen}, {Garc{\'{\i}}a}, \& {Goupil}}]{deheuvels2015}
{Deheuvels}, S., {Ballot}, J., {Beck}, P.~G., {et~al.} 2015, \aap, 580, A96

\bibitem[{{Deheuvels} {et~al.}(2014){Deheuvels}, {Doğan, G.}, {Goupil, M.~J.},
  {Appourchaux, T.}, {Benomar, O.}, {Bruntt, H.}, {Campante, T.~L.},
  {Casagrande, L.}, {Ceillier, T.}, {Davies, G.~R.}, {De Cat, P.}, {Fu, J.~N.},
  {García, R.~A.}, {Lobel, A.}, {Mosser, B.}, {Reese, D.~R.}, {Regulo, C.},
  {Schou, J.}, {Stahn, T.}, {Thygesen, A.~O.}, {Yang, X.~H.}, {Chaplin, W.~J.},
  {Christensen-Dalsgaard, J.}, {Eggenberger, P.}, {Gizon, L.}, {Mathis, S.},
  {Molenda-Żakowicz, J.}, \& {Pinsonneault, M.}}]{deheuvels2014}
{Deheuvels}, S., {Doğan, G.}, {Goupil, M.~J.}, {et~al.} 2014, \aap, 564, A27

\bibitem[{{Deheuvels} {et~al.}(2012){Deheuvels}, {Garc{\'{\i}}a}, {Chaplin},
  {Basu}, {Antia}, {Appourchaux}, {Benomar}, {Davies}, {Elsworth}, {Gizon},
  {Goupil}, {Reese}, {Regulo}, {Schou}, {Stahn}, {Casagrande},
  {Christensen-Dalsgaard}, {Fischer}, {Hekker}, {Kjeldsen}, {Mathur}, {Mosser},
  {Pinsonneault}, {Valenti}, {Christiansen}, {Kinemuchi}, \&
  {Mullally}}]{deheuvels2012}
{Deheuvels}, S., {Garc{\'{\i}}a}, R.~A., {Chaplin}, W.~J., {et~al.} 2012, \apj,
  756, 19

\bibitem[{Di~Mauro {et~al.}(2016)Di~Mauro, Ventura, Cardini, Stello,
  Christensen-Dalsgaard, Dziembowski, Patern{\`o}, Beck, Bloemen, Davies,
  {et~al.}}]{dimauro2016}
Di~Mauro, M., Ventura, R., Cardini, D., {et~al.} 2016, The Astrophysical
  Journal, 817, 65

\bibitem[{{Fuller} {et~al.}(2014){Fuller}, {Lecoanet}, {Cantiello}, \&
  {Brown}}]{fuller2014}
{Fuller}, J., {Lecoanet}, D., {Cantiello}, M., \& {Brown}, B. 2014, ArXiv
  e-prints [\eprint[arXiv]{1409.6835}]

\bibitem[{Goupil {et~al.}(2013)Goupil, Mosser, Marques, Ouazzani, Belkacem,
  Lebreton, \& Samadi}]{goupil2013}
Goupil, M.-J., Mosser, B., Marques, J., {et~al.} 2013, Astronomy \&
  Astrophysics, 549, A75

\bibitem[{Grevesse \& Sauval(1998)}]{grevesse1998}
Grevesse, N. \& Sauval, A. 1998, Space Science Reviews, 85, 161

\bibitem[{{Hekker} {et~al.}(2009){Hekker}, {Kallinger}, {Baudin}, {De Ridder},
  {Barban}, {Carrier}, {Hatzes}, {Weiss}, \& {Baglin}}]{hekker2009}
{Hekker}, S., {Kallinger}, T., {Baudin}, F., {et~al.} 2009, \aap, 506, 465

\bibitem[{{Iglesias} \& {Rogers}(1996)}]{iglesias1996}
{Iglesias}, C.~A. \& {Rogers}, F.~J. 1996, \apj, 464, 943

\bibitem[{Kjeldsen {et~al.}(2008)Kjeldsen, Bedding, \&
  Christensen-Dalsgaard}]{kjeldsen2008}
Kjeldsen, H., Bedding, T.~R., \& Christensen-Dalsgaard, J. 2008, The
  Astrophysical Journal Letters, 683, L175

\bibitem[{Klion \& Quataert(2016)}]{klion2016}
Klion, H. \& Quataert, E. 2016, Monthly Notices of the Royal Astronomical
  Society: Letters

\bibitem[{{Koch} {et~al.}(2010){Koch}, {Borucki}, {Basri}, {Batalha}, {Brown},
  {Caldwell}, {Christensen-Dalsgaard}, {Cochran}, {DeVore}, {Dunham},
  {Gautier}, {Geary}, {Gilliland}, {Gould}, {Jenkins}, {Kondo}, {Latham},
  {Lissauer}, {Marcy}, {Monet}, {Sasselov}, {Boss}, {Brownlee}, {Caldwell},
  {Dupree}, {Howell}, {Kjeldsen}, {Meibom}, {Morrison}, {Owen}, {Reitsema},
  {Tarter}, {Bryson}, {Dotson}, {Gazis}, {Haas}, {Kolodziejczak}, {Rowe}, {Van
  Cleve}, {Allen}, {Chandrasekaran}, {Clarke}, {Li}, {Quintana}, {Tenenbaum},
  {Twicken}, \& {Wu}}]{koch2010}
{Koch}, D.~G., {Borucki}, W.~J., {Basri}, G., {et~al.} 2010, \apjl, 713, L79

\bibitem[{{Kurtz} {et~al.}(2014){Kurtz}, {Saio}, {Takata}, {Shibahashi},
  {Murphy}, \& {Sekii}}]{kurtz2014}
{Kurtz}, D.~W., {Saio}, H., {Takata}, M., {et~al.} 2014, \mnras, 444, 102

\bibitem[{{Ledoux}(1951)}]{ledoux1951}
{Ledoux}, P. 1951, \apj, 114, 373

\bibitem[{{Mosser} {et~al.}(2012){Mosser}, {Goupil}, {Belkacem}, {Marques},
  {Beck}, {Bloemen}, {De Ridder}, {Barban}, {Deheuvels}, {Elsworth}, {Hekker},
  {Kallinger}, {Ouazzani}, {Pinsonneault}, {Samadi}, {Stello}, {Garc{\'{\i}}a},
  {Klaus}, {Li}, {Mathur}, \& {Morris}}]{mosser2012}
{Mosser}, B., {Goupil}, M.~J., {Belkacem}, K., {et~al.} 2012, \aap, 548, A10

\bibitem[{Mosser {et~al.}(2015)Mosser, Vrard, Belkacem, Deheuvels, \&
  Goupil}]{mosser2015}
Mosser, B., Vrard, M., Belkacem, K., Deheuvels, S., \& Goupil, M. 2015,
  Astronomy \& Astrophysics, 584, A50

\bibitem[{{Mosser, B.} {et~al.}(2012){Mosser, B.}, {Goupil, M. J.}, {Belkacem,
  K.}, {Michel, E.}, {Stello, D.}, {Marques, J. P.}, {Elsworth, Y.}, {Barban,
  C.}, {Beck, P. G.}, {Bedding, T. R.}, {De Ridder, J.}, {García, R. A.},
  {Hekker, S.}, {Kallinger, T.}, {Samadi, R.}, {Stumpe, M. C.}, {Barclay, T.},
  \& {Burke, C. J.}}]{mosser2012b}
{Mosser, B.}, {Goupil, M. J.}, {Belkacem, K.}, {et~al.} 2012, \aap, 540, A143

\bibitem[{{Mosser, B.} {et~al.}(2013){Mosser, B.}, {Michel, E.}, {Belkacem,
  K.}, {Goupil, M. J.}, {Baglin, A.}, {Barban, C.}, {Provost, J.}, {Samadi,
  R.}, {Auvergne, M.}, \& {Catala, C.}}]{mosser2013}
{Mosser, B.}, {Michel, E.}, {Belkacem, K.}, {et~al.} 2013, \aap, 550, A126

\bibitem[{Nelder \& Mead(1965)}]{nelder1965}
Nelder, J.~A. \& Mead, R. 1965, The computer journal, 7, 308

\bibitem[{{Paxton} {et~al.}(2011){Paxton}, {Bildsten}, {Dotter}, {Herwig},
  {Lesaffre}, \& {Timmes}}]{paxton2011}
{Paxton}, B., {Bildsten}, L., {Dotter}, A., {et~al.} 2011, \apjs, 192, 3

\bibitem[{{Paxton} {et~al.}(2013){Paxton}, {Cantiello}, {Arras}, {Bildsten},
  {Brown}, {Dotter}, {Mankovich}, {Montgomery}, {Stello}, {Timmes}, \&
  {Townsend}}]{paxton2013}
{Paxton}, B., {Cantiello}, M., {Arras}, P., {et~al.} 2013, \apjs, 208, 4

\bibitem[{{Paxton} {et~al.}(2015){Paxton}, {Marchant}, {Schwab}, {Bauer},
  {Bildsten}, {Cantiello}, {Dessart}, {Farmer}, {Hu}, {Langer}, {Townsend},
  {Townsley}, \& {Timmes}}]{paxton2015}
{Paxton}, B., {Marchant}, P., {Schwab}, J., {et~al.} 2015, \apjs, 220, 15

\bibitem[{{P{\'e}rez Hern{\'a}ndez} {et~al.}(2016){P{\'e}rez Hern{\'a}ndez},
  {Garc{\'{\i}}a}, {Corsaro}, {Triana}, \& {De Ridder}}]{hernandez2016}
{P{\'e}rez Hern{\'a}ndez}, F., {Garc{\'{\i}}a}, R.~A., {Corsaro}, E., {Triana},
  S.~A., \& {De Ridder}, J. 2016, \aap, 591, A99

\bibitem[{{Pijpers} \& {Thompson}(1994)}]{pijpers1994}
{Pijpers}, F.~P. \& {Thompson}, M.~J. 1994, \aap, 281, 231

\bibitem[{{Rogers}(2015)}]{rogers2015}
{Rogers}, T.~M. 2015, \apjl, 815, L30

\bibitem[{{Saio} {et~al.}(2015){Saio}, {Kurtz}, {Takata}, {Shibahashi},
  {Murphy}, {Sekii}, \& {Bedding}}]{saio2015}
{Saio}, H., {Kurtz}, D.~W., {Takata}, M., {et~al.} 2015, \mnras, 447, 3264

\bibitem[{{Shibahashi}(1979)}]{shibahashi1979}
{Shibahashi}, H. 1979, \pasj, 31, 87

\bibitem[{{Stello} {et~al.}(2009){Stello}, {Chaplin}, {Bruntt}, {Creevey},
  {Garc{\'{\i}}a-Hern{\'a}ndez}, {Monteiro}, {Moya}, {Quirion}, {Sousa},
  {Su{\'a}rez}, {Appourchaux}, {Arentoft}, {Ballot}, {Bedding},
  {Christensen-Dalsgaard}, {Elsworth}, {Fletcher}, {Garc{\'{\i}}a}, {Houdek},
  {Jim{\'e}nez-Reyes}, {Kjeldsen}, {New}, {R{\'e}gulo}, {Salabert}, \&
  {Toutain}}]{stello2009}
{Stello}, D., {Chaplin}, W.~J., {Bruntt}, H., {et~al.} 2009, \apj, 700, 1589

\bibitem[{{Townsend} \& {Teitler}(2013)}]{townsend2013}
{Townsend}, R.~H.~D. \& {Teitler}, S.~A. 2013, \mnras, 435, 3406

\bibitem[{Triana {et~al.}(2015)Triana, Moravveji, Pápics, Aerts, Kawaler, \&
  Christensen-Dalsgaard}]{triana2015}
Triana, S.~A., Moravveji, E., Pápics, P.~I., {et~al.} 2015, The Astrophysical
  Journal, 810, 16

\bibitem[{Unno(1989)}]{unno1989}
Unno, W. 1989, Nonradial oscillations of stars (University of Tokyo Press)

\end{thebibliography}


\begin{appendix}

\section{Hare-and-hounds tests}
\label{app}
The adequacy of the method we used to obtain the seismic models
can be assessed by a hare-and-hounds exercise. First, we selected
the best seismic model of KIC007619745 as our `true' model, then we
assumed a known internal rotation profile and derived
`observations' by adding some noise to the `true' mode frequencies.
The errors are assumed to be the same as in the actual observations.
Then we used our grid + downhill simplex search strategy to obtain
a seismic model that best reproduces these `observations'.

We have deliberately used a different mixing length parameter
($\alpha_{\mathrm{MLT}}=1.7$)
and a different atmospheric correction power $b$ during the search ($b=4.9$)
compared to the true model (which has $\alpha_{\mathrm{MLT}}=1.9$, $b=4.81$),
thus further guaranteeing that the best model found is not identical to the
true model. The best model properly recovers both the initial metallicity [Fe/H] 
and the initial helium content $Y$. The inferences on the internal rotation
averages are also in good agreement with the true profile
within $2\sigma$. However, the same cannot be said of other stellar parameters such as mass, $f_{\mathrm{ov}}$ , or radius.

\begin{figure}
\begin{center}
\includegraphics[width=1.0\linewidth]{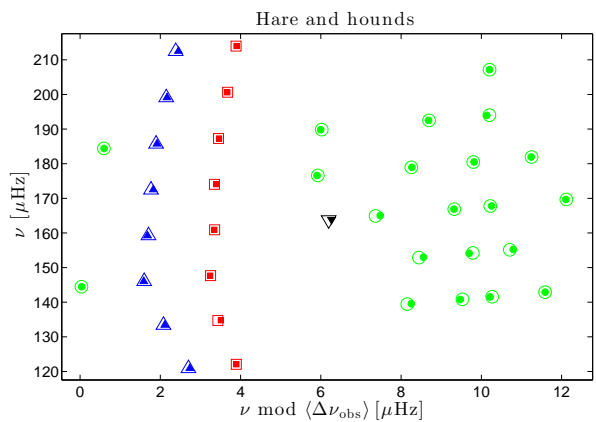}
\caption{\'Echelle diagram comparing the noise-added true model frequencies
(hollow symbols) and the
best model found by a grid + simplex search approach (filled symbols).
Red squares correspond 
to $l=0$ modes, green circles to $l=1$, blue triangles to $l=2,$ and inverted 
black triangles to $l=3$.}
\label{ech2}
\end{center}
\end{figure}

An \'echelle diagram comparing the frequencies of the best model with
the noise-added true model is shown in Fig.\,\ref{ech2}. Table \ref{hah_tab} lists other stellar
parameters from the two models.
{
The results for the internal rotation rates using different methods
based on this best model are displayed in
Fig.\,\ref{omg2} (bottom, right panel). Here we used
a `true' rotation profile according to Eq.~\ref{klionprof} with
$\Omega_c=0.7~\mu$Hz, $\Omega_m=0.4~\mu$Hz, and $\Omega_e=0.1\mu~$Hz. Compare also
with the top and middle panel in Fig.~\ref{testnonoise}.
}
\begin{table}[h]
\caption{Hare-and-hounds model comparison}             
\label{hah_tab}      
\centering                          
\begin{tabular}{c c c c c c}        
\hline\hline                 
Model & $M/M_{\odot}$ & $R/R_{\odot}$ & $\Delta\nu$ [$\mu$Hz] & $T_\mathrm{eff}$ [K] & $f_{\mathrm{ov}}$ \\    
\hline                        
 true & 1.34          & 5.12          & 13.28                &  4969                & 0.0183            \\      
 best & 1.46          & 5.27          & 13.26                &  4872                & 0.0057            \\ 
\hline                                   
\end{tabular}
\end{table}

\begin{figure*}[h!]
\begin{center}$
\begin{array}{cc}
\includegraphics[width=0.45\linewidth]{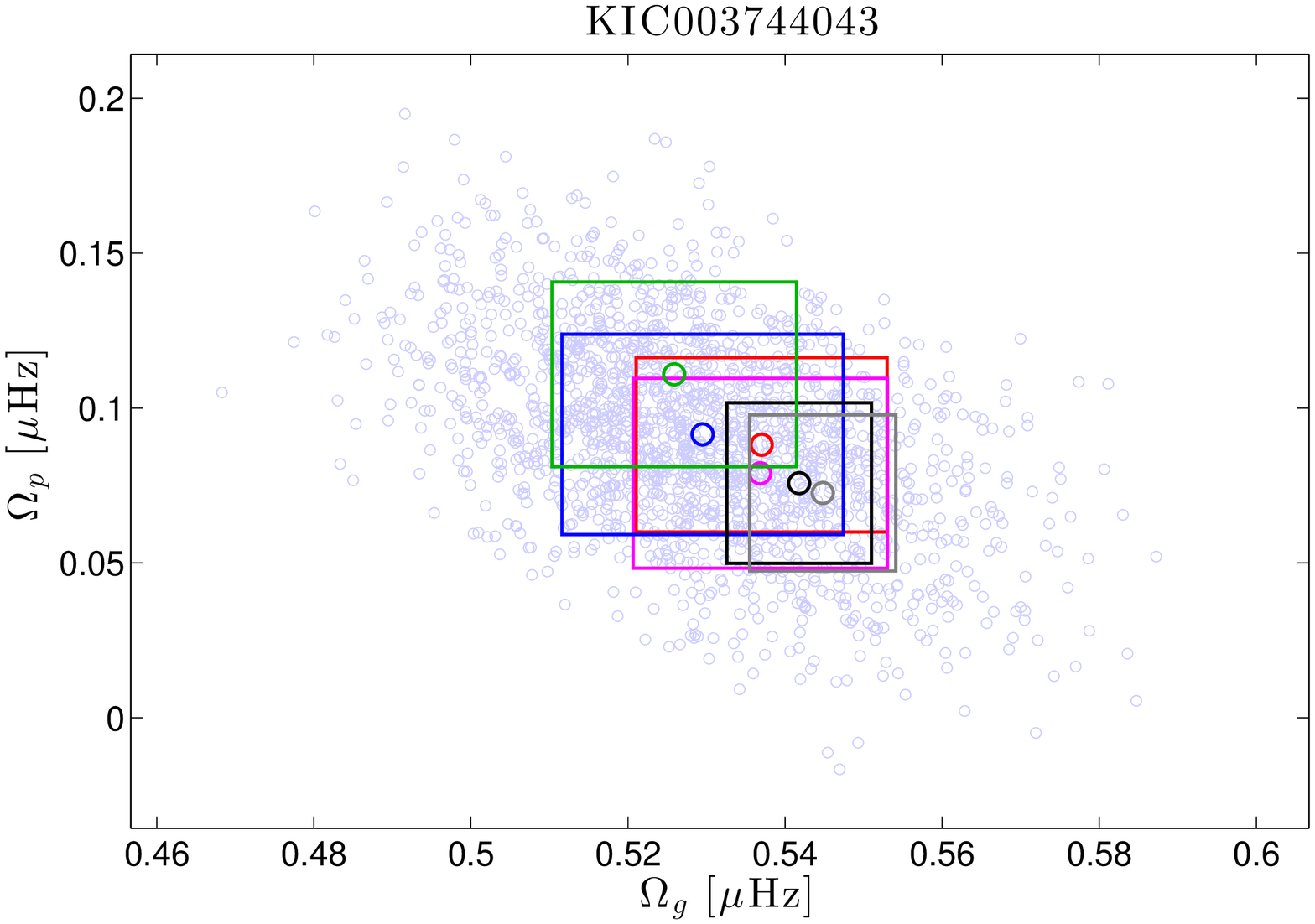}&
\includegraphics[width=0.45\linewidth]{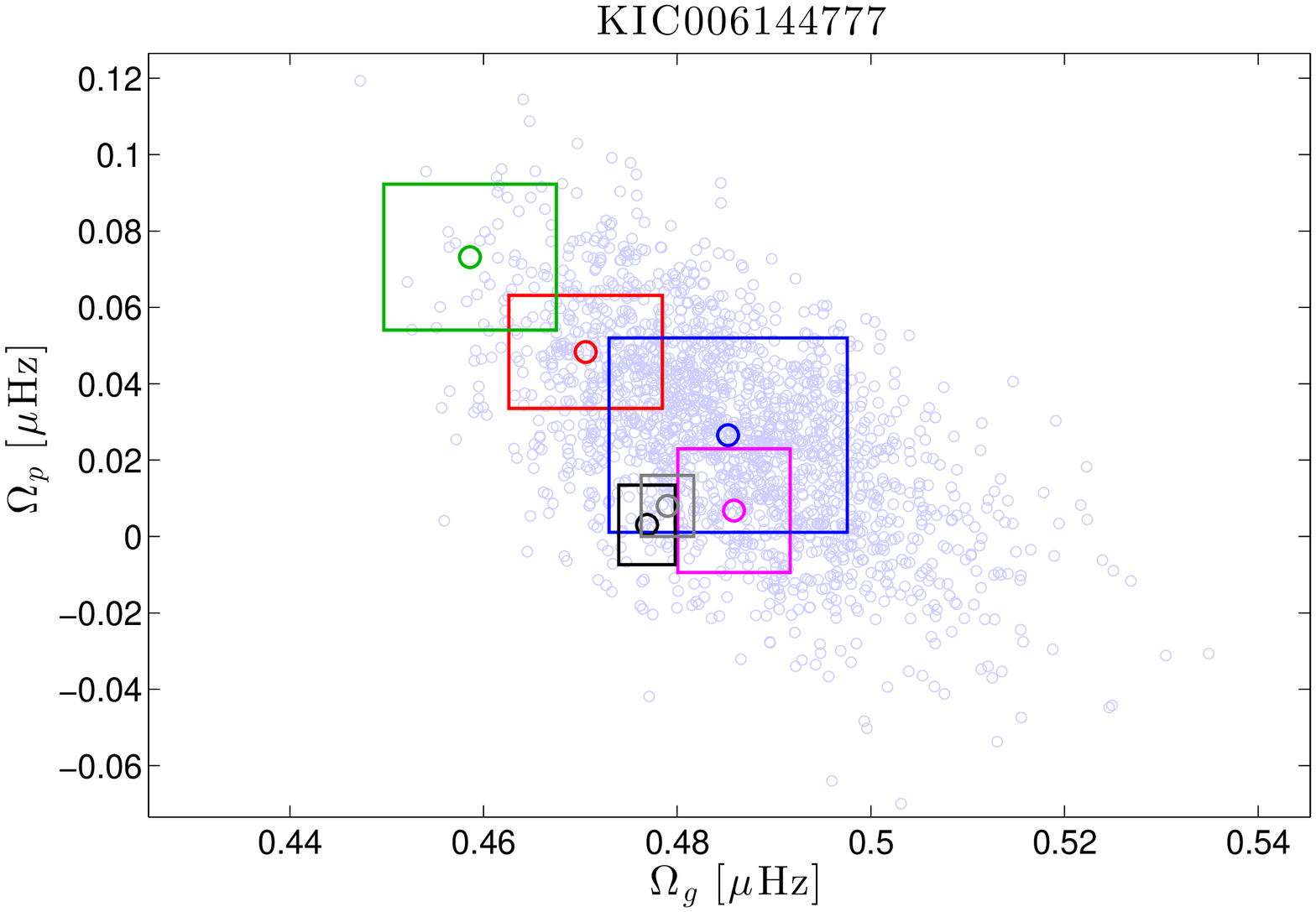}\\
\includegraphics[width=0.45\linewidth]{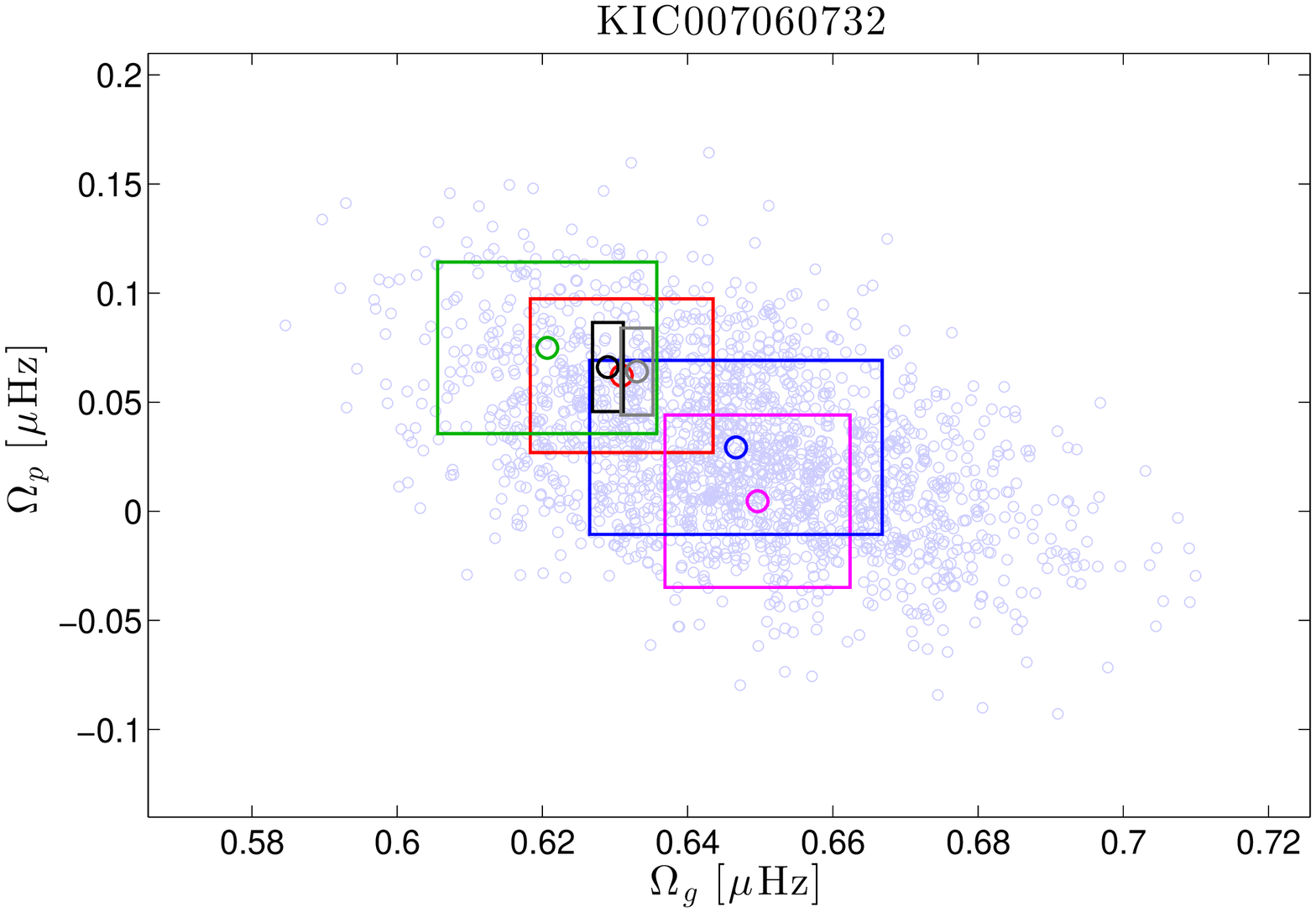}&
\includegraphics[width=0.45\linewidth]{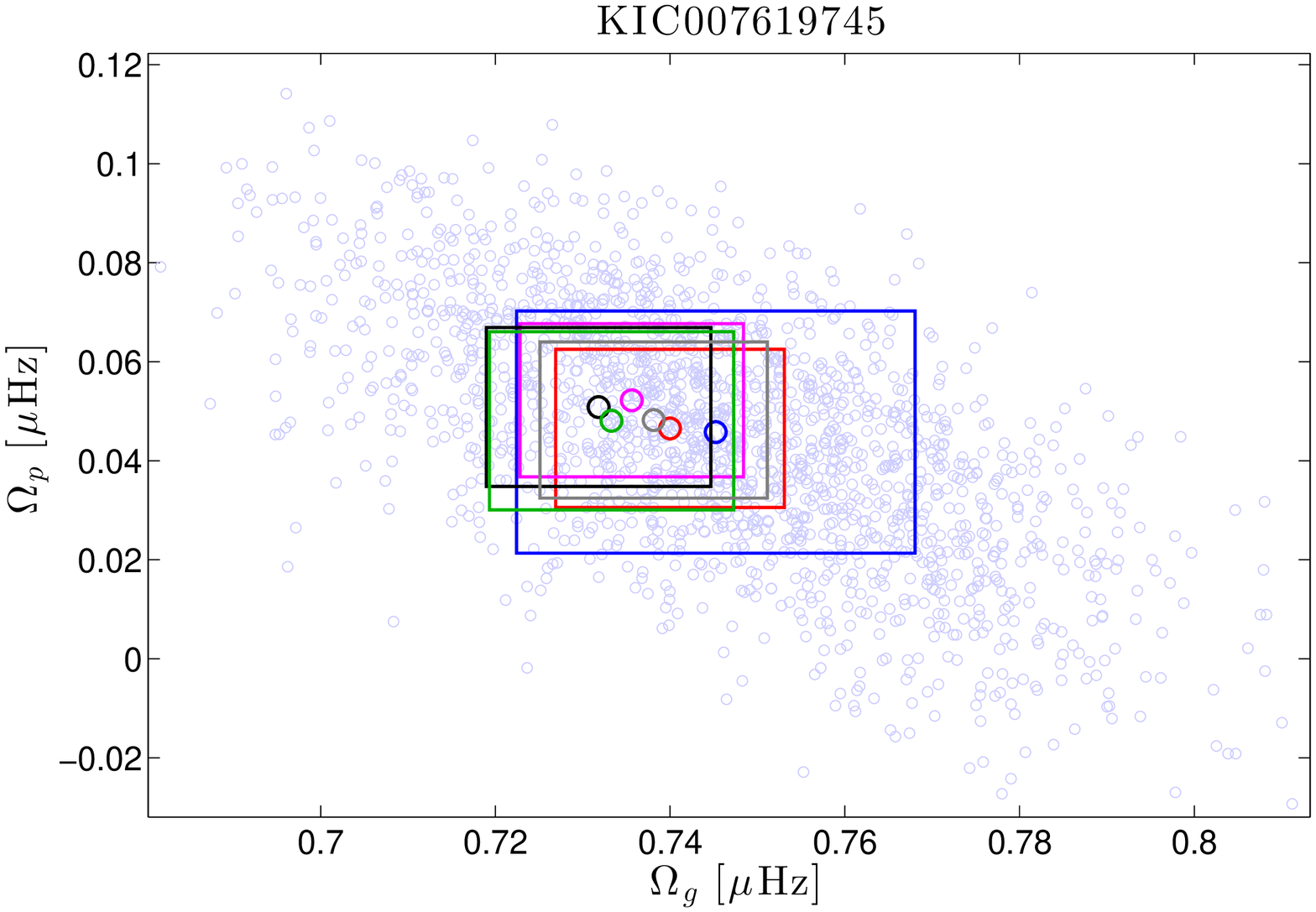}\\
\includegraphics[width=0.45\linewidth]{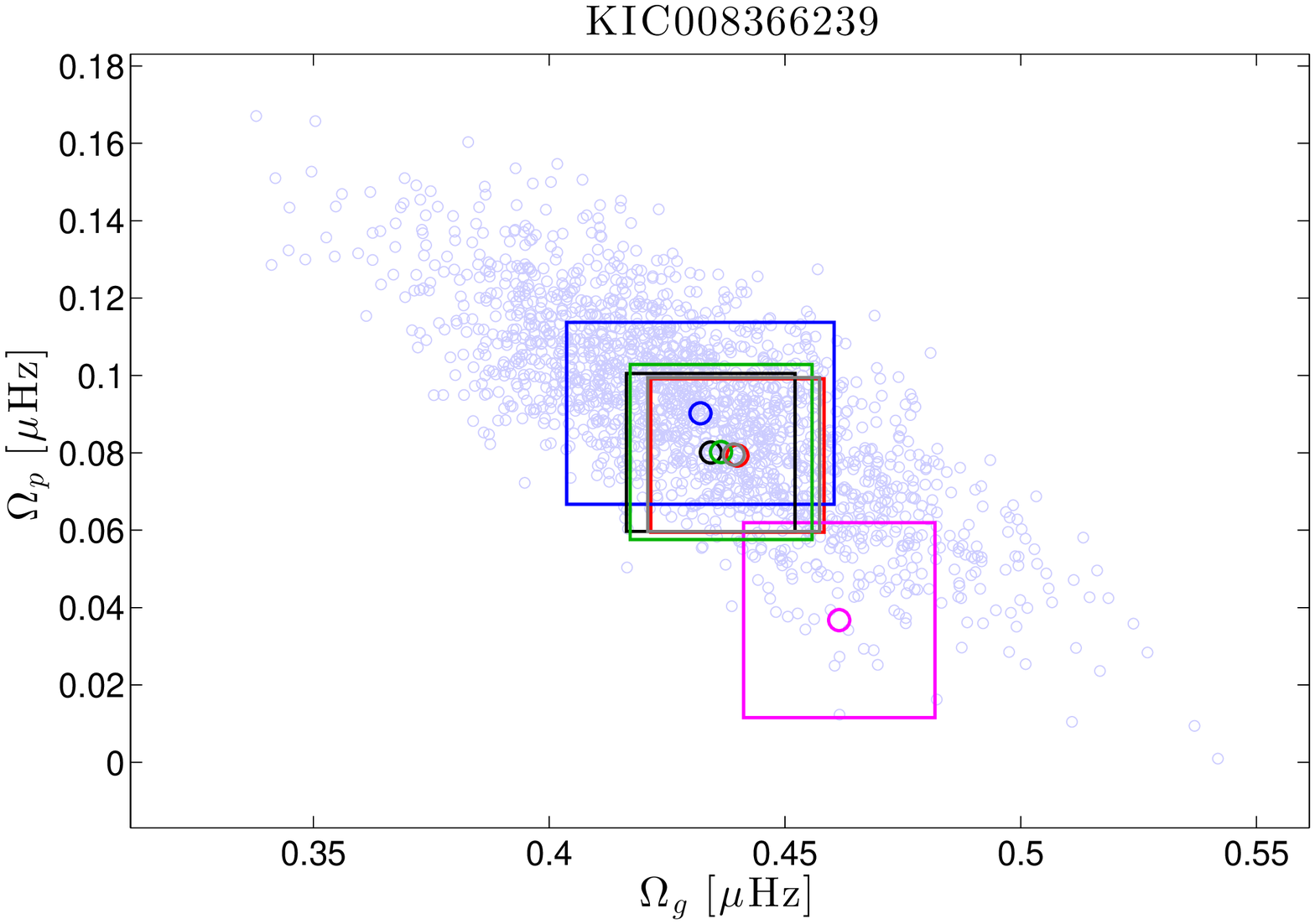}&
\includegraphics[width=0.45\linewidth]{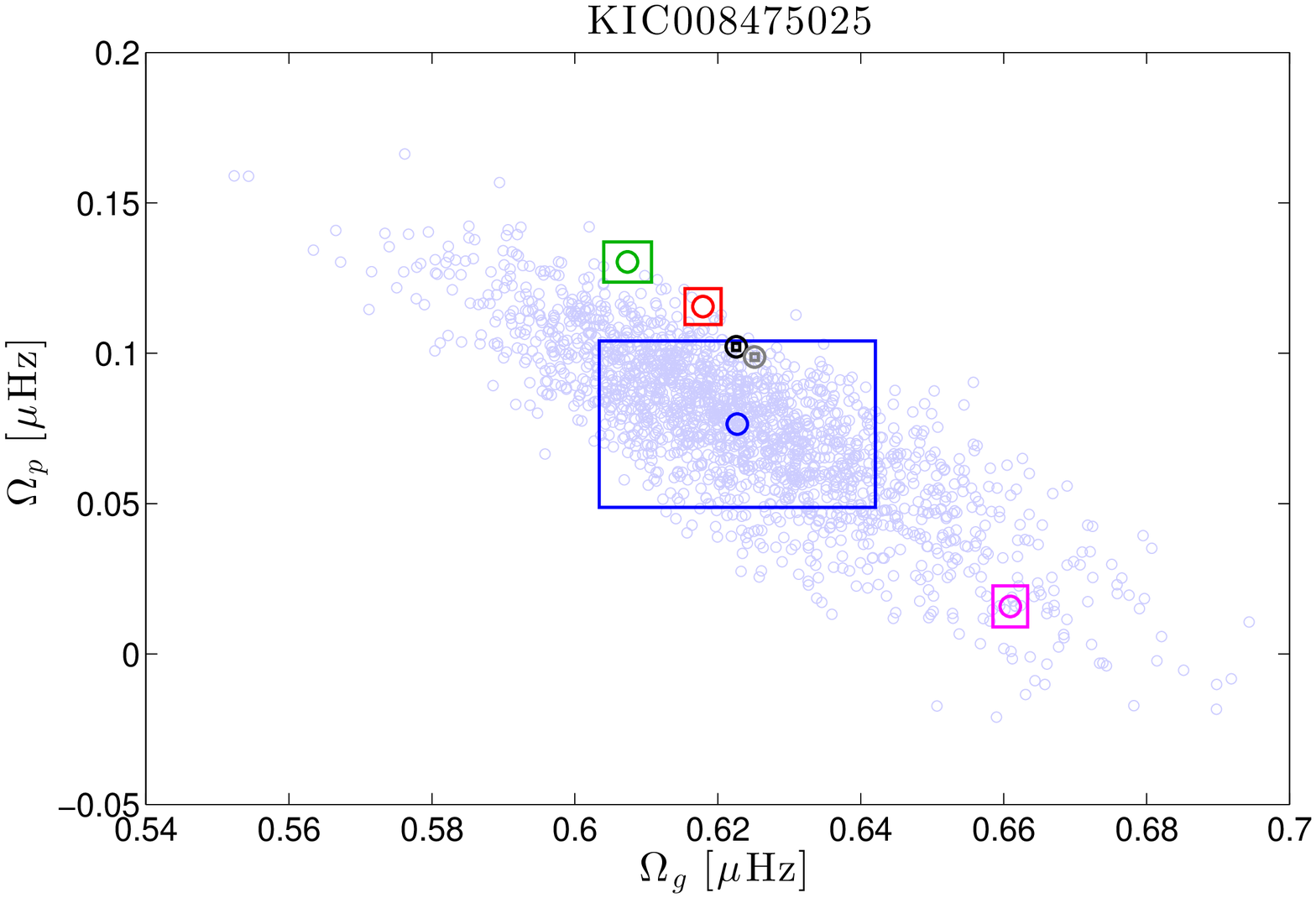}\\
\includegraphics[width=0.45\linewidth]{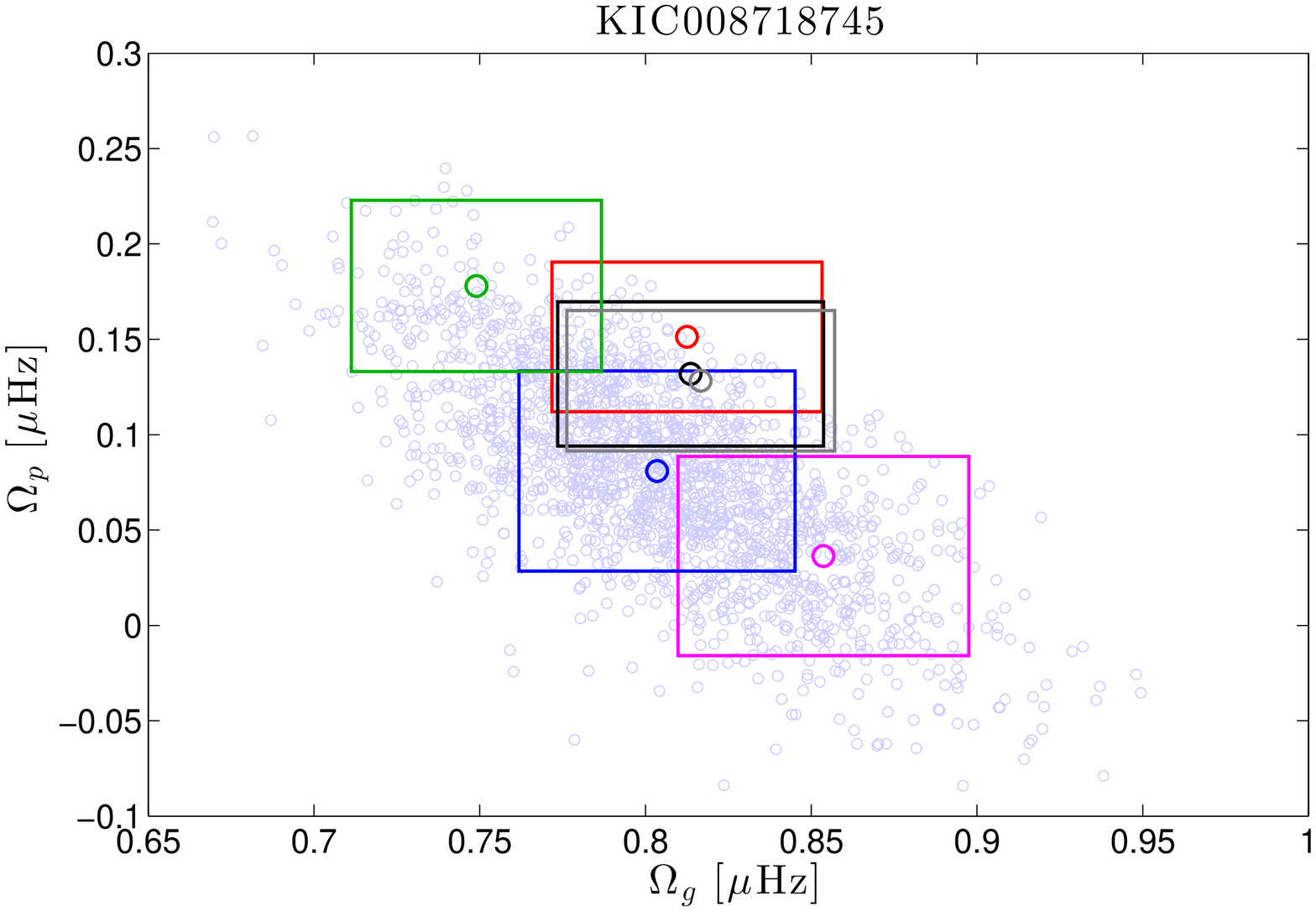}&
\includegraphics[width=0.45\linewidth]{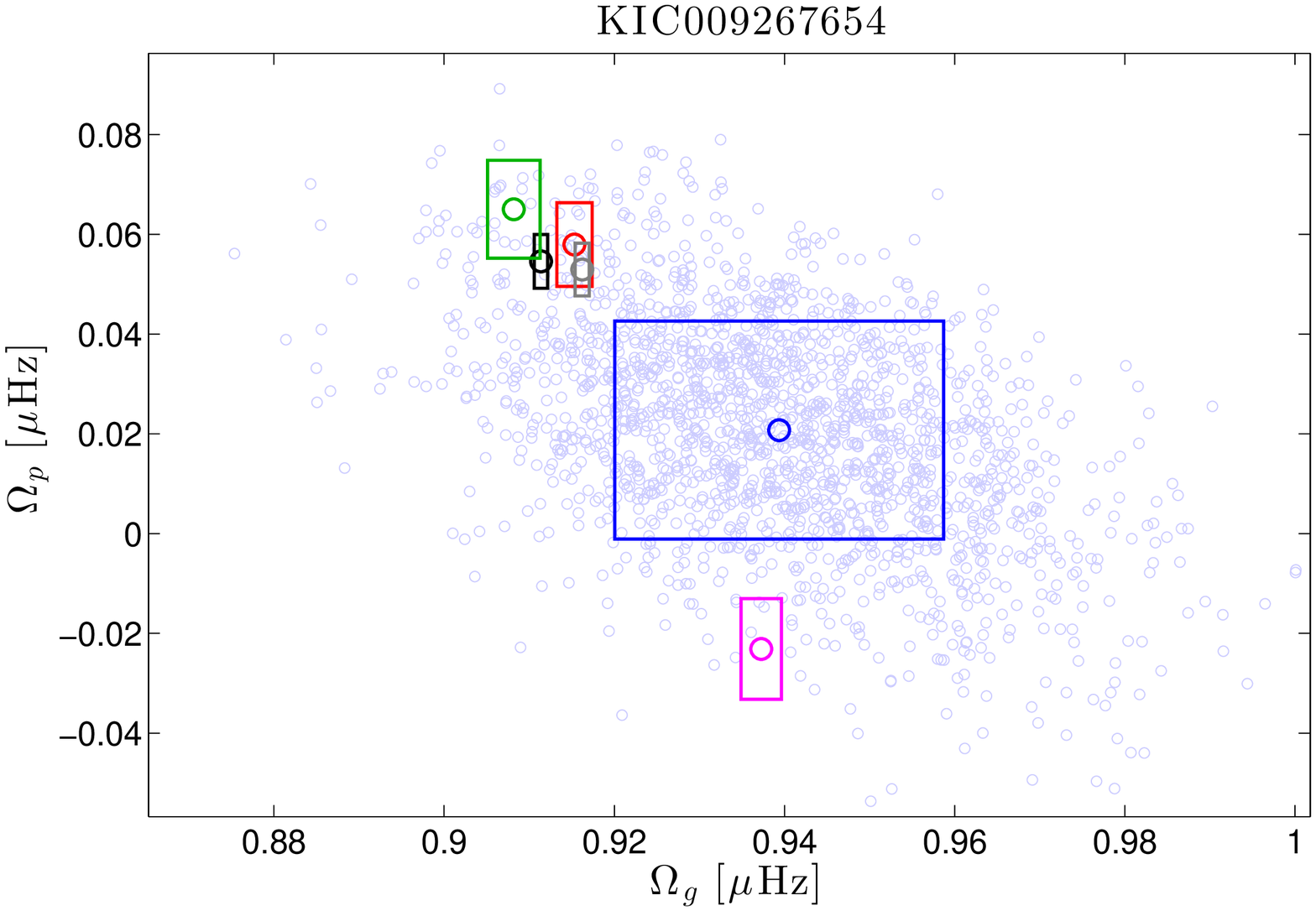}\\
\end{array}$
\end{center}
\caption{Rotation rates for all targets. Rotation
rate of the envelopes ($\Omega_p$) on the ordinates, rotation rate of the cores ($\Omega_g$)
on the abscissae.
We show the results inferred from two-zone inversions (black), from
SOLA inversions (green), from linear fits of $\delta$ vs. $\zeta_\mathrm{mod}$ (red), 
from linear fits of $\delta$ vs. $\zeta_\mathrm{as}$ (magenta), and from Bayesian inference (gray),
 all with their corresponding 1$\sigma$ error margins. The cloud of light blue circles
 is the result of applying linear fits of $\delta$ vs. $\delta P/\Delta \Pi_1$
  sampling randomly from the splittings $\delta$ and the observed mode 
  frequencies $\nu$, assuming a normal distribution for both. The blue error box indicates the
  $1\sigma$ standard deviation of the resulting distribution on each axis.}
\label{omg1}
\end{figure*}

\begin{figure*}
\begin{center}$
\begin{array}{cc}
\includegraphics[width=0.45\linewidth]{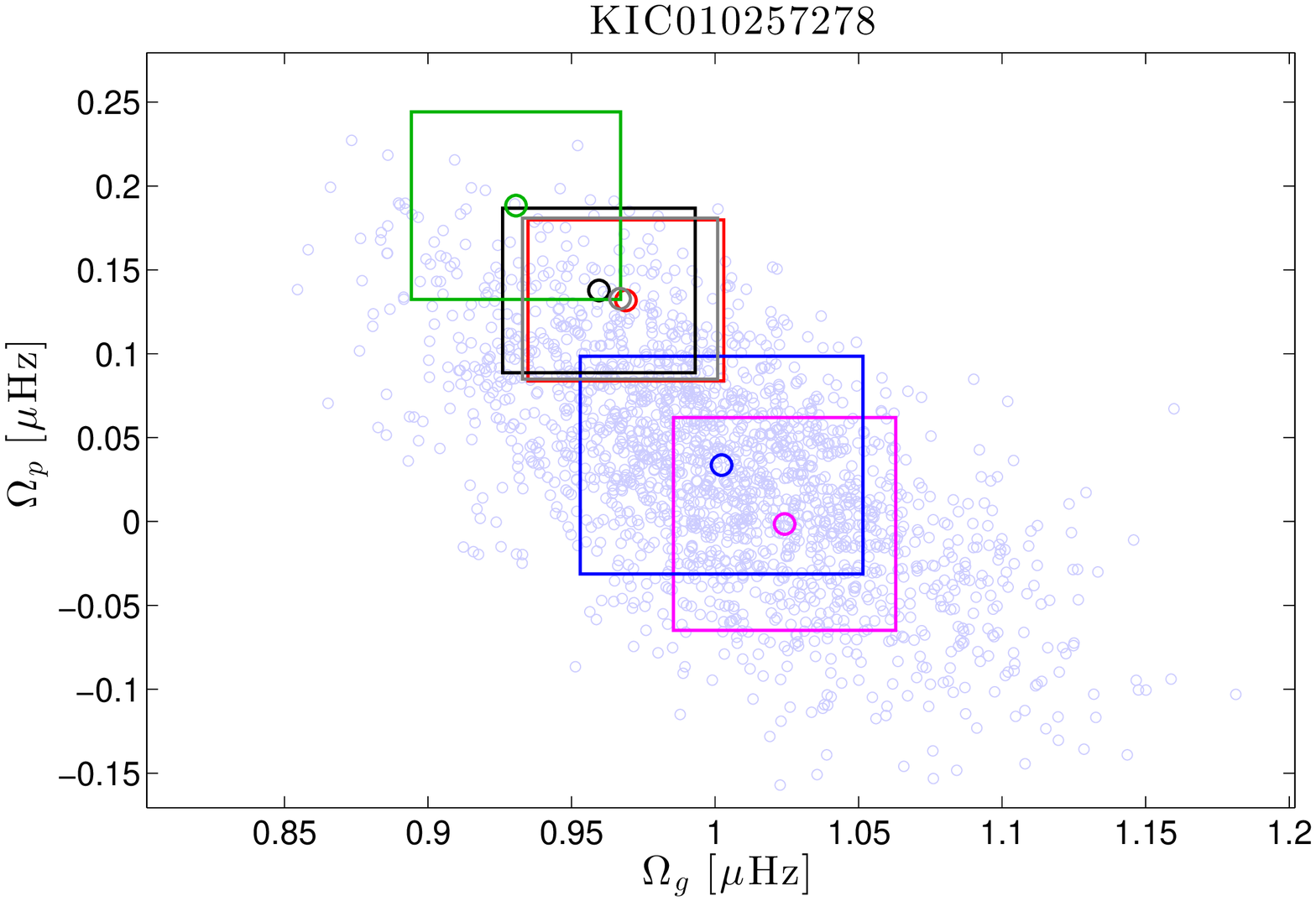}&
\includegraphics[width=0.45\linewidth]{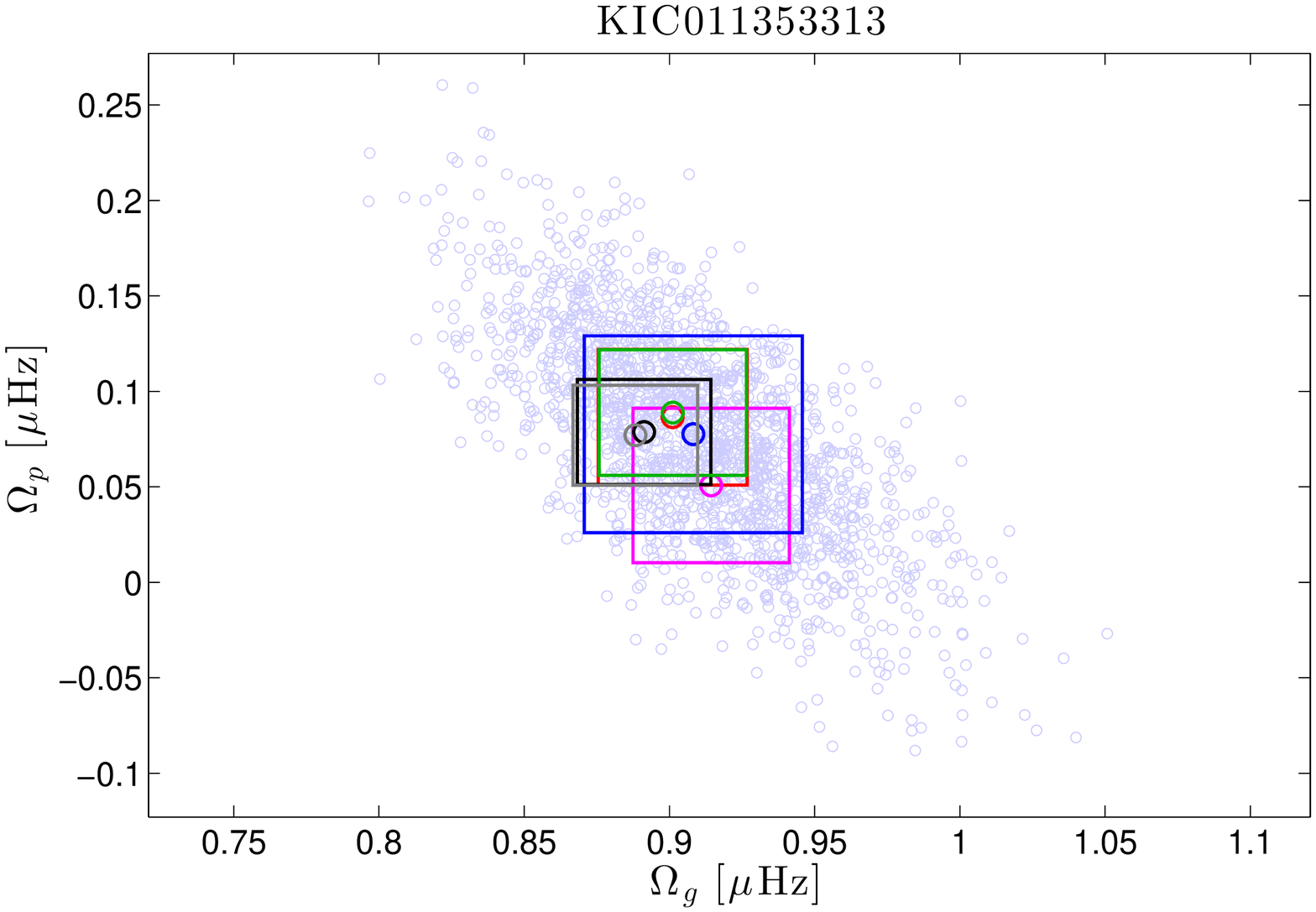}\\
\includegraphics[width=0.45\linewidth]{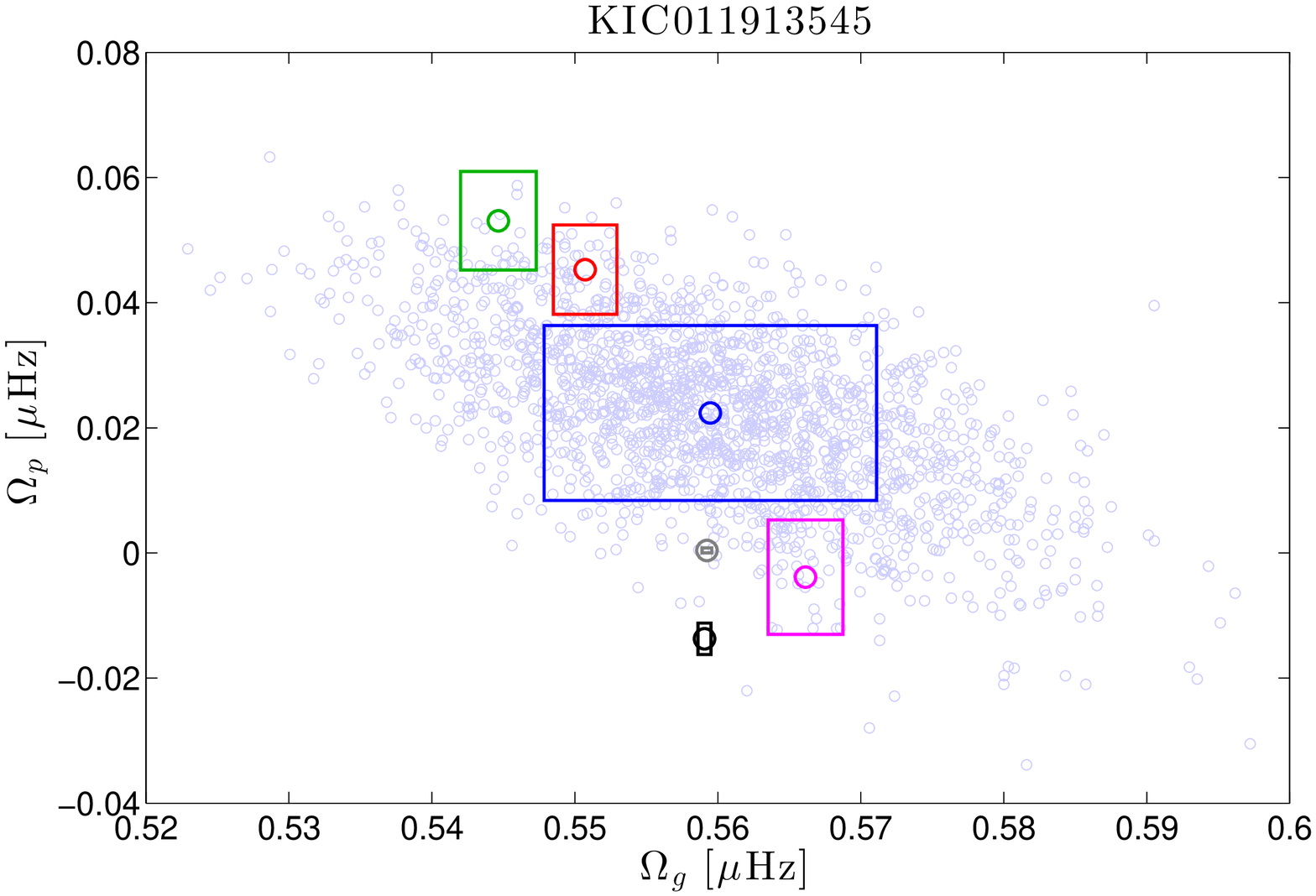}&
\includegraphics[width=0.45\linewidth]{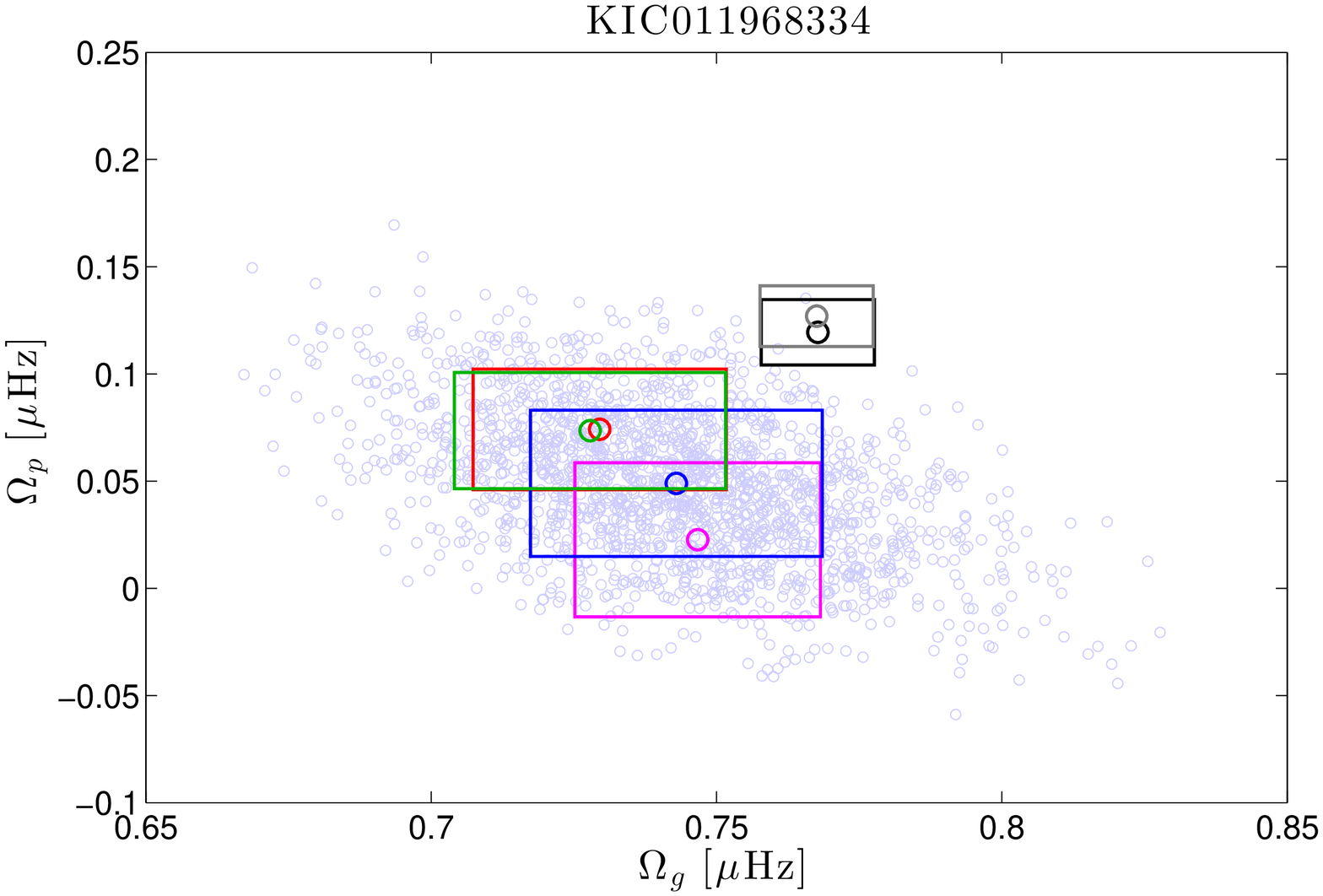}\\
\includegraphics[width=0.45\linewidth]{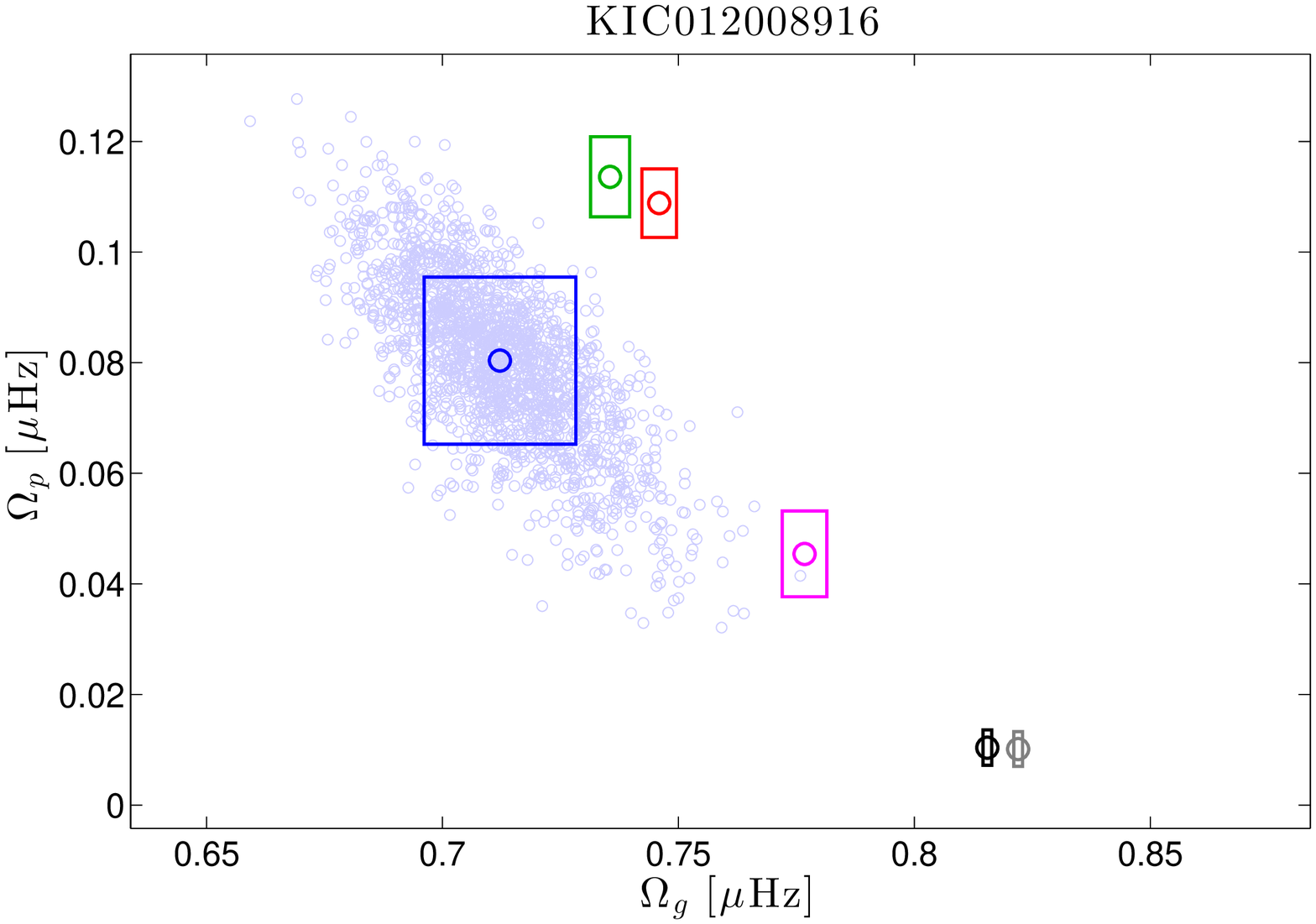}&
\includegraphics[width=0.45\linewidth]{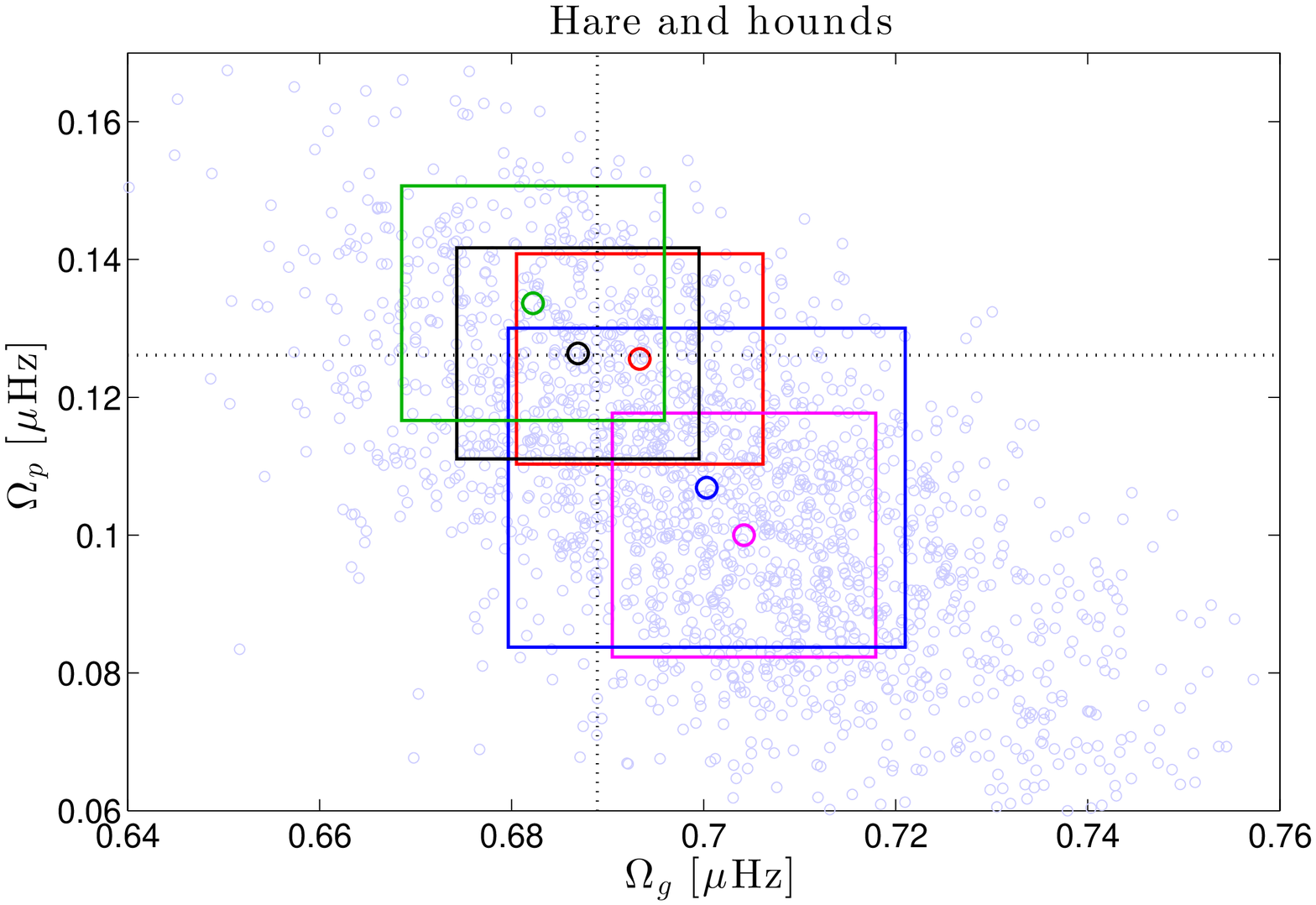}\\

\end{array}$
\end{center}
\caption{Continuation of Fig.\,\ref{omg1}. Rotation
rate of the envelopes ($\Omega_p$) on the ordinates, rotation rate of the cores ($\Omega_g$)
on the abscissae.
We show the results inferred from two-zone inversions (black), from
SOLA inversions (green), from linear fits of $\delta$ vs. $\zeta_\mathrm{mod}$ (red), 
from linear fits of $\delta$ vs. $\zeta_\mathrm{as}$ (magenta), and from Bayesian inference (gray),
 all with their corresponding 1$\sigma$ error margins. The cloud of light blue circles
 is the result of applying linear fits of $\delta$ vs. $\delta P/\Delta \Pi_1$
  sampling randomly from the splittings $\delta$ and the observed mode 
  frequencies $\nu$, assuming a normal distribution for both. The blue error box indicates the
  $1\sigma$ standard deviation of the resulting distribution on each axis.}
\label{omg2}
\end{figure*}

\end{appendix}

\end{document}